\documentclass{article}
\usepackage[utf8]{inputenc}

\usepackage{cite}
\usepackage{amsmath,amssymb,amsfonts}
\usepackage{algorithmic}
\usepackage{enumerate}
\usepackage{graphicx}
\usepackage{graphics}
\usepackage{url}
\usepackage{hyperref}
\usepackage{textcomp}
\usepackage{mathtools, cuted}
\usepackage{stfloats}
\usepackage{multirow}
\usepackage{subcaption}

\newtheorem{lem}{Lemma}

\newtheorem{cor}{Corollary}
\newtheorem{defn}{Definition}
\newtheorem{thm}{Theorem}

\newcounter{example}[section]

\title{A Unified Framework  for Consensus and Synchronization on Lie Groups admitting a Bi-Invariant Metric}
\author{Rama Seshan, Ravi N Banavar, Arun D Mahindrakar }
\date{September 2022}

\begin{document}

\maketitle

\begin{abstract}           For a finite number of agents evolving on a Euclidean space and linked to each other by a connected graph, the Laplacian flow that is based on the inter-agent errors, ensures consensus or synchronization for both first and second-order dynamics.  When such agents evolve on a circle (the Kuramoto oscillator), the flow that depends on the sinusoid of the inter-agent error angles generalizes the same. In this work, it is shown that the Laplacian flow and the Kuramoto oscillator are special cases of a more general theory of consensus on Lie groups that admit  bi-invariant metrics. Such a theory not only enables generalization of these consensus and synchronization algorithms to Lie groups but also provide insight on to the abstract group theoretic and differential geometric properties that ensures convergence in Euclidean space and the circle.

% Abstract of not more than 200 words.

\end{abstract}

%\end{frontmatter}

\section{Introduction}

There is a huge scope for the design and analysis of algorithms that coordinate a system of identical autonomous agents - examples include formation of vehicles \cite{1205192}, oscillator synchronization \cite{STROGATZ20001}, spacecraft formations \cite{4777127}, study of flocking mechanisms in living beings such as birds or fish \cite{1582620}, mobile sensor networks, social dynamics \cite{10.2307/2285509} and mechanical system networks in robotics \cite{HANMANN2006176}.

For agents that evolve on Euclidean spaces of any finite dimension, design and analysis of consensus algorithms, both in continuous and discrete time exists in the literature and is well established \cite{BulloNS}. The adjacency matrix that describes the communication topology of the connectivity of the network of agents is used in generating a linear flow map on the configuration space of the multi-agent system (which is a direct product of the configuration space of each agent), called the Laplacian flow in continuous-time. The Laplacian flow is designed such that the state of the system, starting from any initial condition, converges asymptotically ( exponentially for linear systems) to the consensus subspace where all agents share the same state. In continuous time, first-order and second-order Laplacian flow on Euclidean spaces serve as standard algorithms for consensus for identical agents evolving on Euclidean spaces, as extensively analyzed  in \cite{BulloNS}. They are recalled in a future section. The central theme is that each agent has a velocity or acceleration that is directly proportional to a weighted sum of its displacement from each of its neighbors.  So, a spring like force is created where each agent is pulled towards the position of all of its neighbors. This forces them to eventually attain consensus.

We refer to agents evolving on the circle as oscillators. Oscillator phase consensus and velocity synchronization has also been extensively  observed, analyzed and studied ever since the observation by Huygens \cite{Huy} that two pendulum clocks hung in the same frame eventually go exactly $180^{\circ}$ out of step. Excellent surveys about oscillator synchronization can be found in  \cite{STROGATZ20001} , \cite{BulloNS}   and \cite{DORFLER20141539}  . Once again, the  theme is the same. Each agent has two components. A constant natural angular velocity with which it wants to maintain and, an angular velocity  that is directly proportional to a weighted sum of the sines of its angular separation from each of its neighbors.  So, a spring like force is created where each agent is pulled towards the angle of all of its neighbors. This forces them to eventually attain consensus. In case of second-order agents with no intrinsic preference for any angular velocity, the consensus term alone appears as an acceleration along with a dissipative damping. 

One notes that both the Euclidean space and the unit circle are Lie groups where one can sensibly define an inter-agent error in terms of the difference in linear or angular displacement between two agents. So, one is tempted to ask if these results can be unified and extended to a general theory of consensus on Lie groups, where one can define a left (or right) invariant error between two agents, whose deviation from identity, captures the degree of separation between them.  There have been many attempts to generalize this theory of consensus and formations to such non-Euclidean, non-circular spaces \cite{SEPULCHRE201156}. For vehicular and rigid body systems, the consensus happens on $SO(n),SE(n)$, where $n=2,3$. There is a huge body of work that deals with consensus on each of these individual spaces separately. However, a common and elegant geometric framework to capture all these theories under a single umbrella is wanting, not only for mathematical completeness and elegance but also to give more insights into the nature of the problem. 

There are also generalizations of consensus to Riemannian manifolds \cite{rm} but these turn out to lack any elegant structure and involve intense computation of the parallel transport map which is not just involved but is also not globally defined. Lie groups provide a nice framework that are neither too general like Riemannian manifolds nor too special like Euclidean spaces or tori. They are generic enough to cover a wide range of systems, especially aerospace and mechanical systems, but also have enough structure to generalize the geometric insights of the Laplacian flow and oscillator synchronization directly. In \cite{doi:10.1137/060673400}, consensus on compact Lie groups is studied using a generalization of the mean to Lie groups. A unified theory for consensus and coordinated motion in Lie groups is presented in \cite{sarlette2010}, 
%but this addresses first-order agent dynamics, where velocities are assumed to be the control input. But all mechanical agents are  essentially second-order in nature, and restricting them top first-order would be an unrealistic assumption. The configuration space of many mechanical systems in
%aerospace and mechanical engineering is a Lie group, which also come equipped with a natural choice of a left/right invariant metric that signifies the inertial property of the agent. Indeed, in \cite{5404366} itself, it is claimed that its algorithm can be used as a high-level controller or as a preliminary step towards an integrated mechanical controller. For Euclidean spaces, the generalization from first-order consensus algorithms to a second-order consensus algorithm is trivial as the tangent space at any point can be identified naturally with the configuration space itself. But this is not so for non-Euclidean spaces. 
but this  does not handle first-order consensus, second-order consensus and velocity synchronization together in a unified framework, and insights into tangible mechanical interpretations, such as being driven by a spring-like potential energy between agents, are absent. 

These developments prompted the authors of this article to  develop a theory that directly builds on the Laplacian flow and coupled oscillator theory on Lie groups. A Lie group with a bi-invariant metric is considered and a preliminary generalization of Laplacian flow to such Lie groups was done in \cite{rama-ecc2022}. Here, the authors used gradients of a special class of potential functions on Lie groups, and the left-invariant inter-agent errors to directly generalize the second-order Laplacian flow and oscillator consensus to Lie groups with a bi-invariant metric. But the equilibrium analysis and stability convergence was only presented for a special communication structure - the line graph - where an agent $i$ is connected to just agents $i-1$ and $i+1$. The present work is built upon that previous preliminary work and  contains the following
significant additions:

\begin{enumerate}
    \item A unified geometric theory for first-order and second-order consensus, and velocity synchronization  on Lie groups admitting a bi-invariant metric is presented and the stability analysis of all of them are carried in a united manner, using special classes of potential functions on Lie groups (named \textbf{$G$-Polar Morse functions}) that mimic the potential energy of a spring and also which are compatible with the group structure.
    \item The consensus equilibrium for first and second-order consensus is proven to be locally exponentially stable for any undirected, weighted, connected graph.
    \item For connected trees, the consensus equilibrium is shown to be asymptotically stable with an almost global domain of attraction.
    \item A necessary condition for existence of synchronous solution is presented for velocity synchronization dynamics, when each agent has a non-uniform natural velocity.
    \item Sufficient conditions for the existence of velocity synchronous solutions are presented for various special graphs like stars, line graphs and trees.
    \item Velocity synchronous solutions are guaranteed to exist when the difference in the natural velocities of the agents are sufficiently small in magnitude.
    \item Any synchronization equilibrium with inter-agent errors at steady-state that are in some neighborhood $U$ (explicitly characterized in the following sections) of identity , is shown to be locally exponentially stable for any connected graph.
\end{enumerate}
\section{Mathematical Preliminaries}
Let $\mathbb{R}$ denote the set of real numbers, $\mathbb{R}_+$, the set of non-negative real numbers. The cardinality of a finite set $A$ is denoted by $|A|$.  Let ($V,\langle.\rangle$) be an inner product space. Denote by $||v||$, the norm of a vector $v$ induced by the inner product. Let $T:V\rightarrow V$ be a linear operator. The operator $T$ is said to be symmetric if it satisfies for all $v,w\in V$, $\langle w,Tv\rangle=\langle v, Tw \rangle$ and is said to be skew-symmetric if it satisfies for all $v,w\in V$, $\langle w,Tv\rangle=-\langle v, Tw \rangle$. Then, we have the following result. It then follows that the spectrum (set of eigen values of an operator) of a symmetric operator consists of real numbers and the spectrum of a skew-symmetric linear operator consists of purely imaginary numbers. An important result from linear algebra is reviewed.
\begin{thm} \cite{kunzz}
\label{basic}
  Let $S$ be a linear transformation in an inner product space such that $S=S_{sym}+S_{sks}$ where $S_{sym}$ is a symmetric operator and $S_{sks}$ is a skew-symmetric operator. Then, $S$ has its spectrum in the  closed (or respectively open) right-half of the complex plane  if and only if $S_{sym}$ is positive semi-definite  (or respectively positive-definite)  - that is, all of the real eigen values of $S_{sym}$ are non-negative (or respectively positive).
\end{thm}
That is, the stability of a dynamical system can be inferred by just investigating the stability of its symmetric part. The addition of a skew-symmetric matrix to a stable symmetric matrix does not affect its stability. The spectrum still remains in the left-half of the complex plane if it originally were so, to begin with.

As a Corollary, noting that a matrix with all of its eigen-values  in open right-half of the complex plane has full rank, we note that adding a skew-symmetric matrix to a symmetric positive-definite matrix, maintains its full rank.

\subsection{Graph Theoretic Preliminaries}

A weighted undirected graph $\mathcal{G}$, is defined as a finite set of vertices , denoted by $\mathcal{V}=\{1,2,\cdots,n\}$, and a weight function $\mathcal{W}:\mathcal{V}\times\mathcal{V}\rightarrow \mathbb{R}_+$ such that
\begin{itemize}
    \item $\mathcal{W}$ is symmetric, that is, $\mathcal{W}(i,j)=\mathcal{W}(j,i) \hspace{3mm}\forall \hspace{1mm} i,j$. 
    \item $\mathcal{W}_{ii}=0$.
\end{itemize}
The symmetric matrix $A$ whose entries are defined as $A=[a_{ij}]:=[\mathcal{W}(i,j)]$ is called the adjacency matrix. The degree of a node $i$ is defined as $deg(i)=\sum_{j\neq i} a_{ij}$.
% A symmetric matrix is said to be stochastic if all of its row (or column) sums is 1. i.e. $\sum_{j=1}^na_{ij}=1$. 
From now on, we will assume we have a weighted undirected symmetric graph whose adjacency matrix is symmetric
% and stochastic
.

The following basic definitions from graph theory are recalled:
\begin{itemize}
    \item A distinct pair of vertices $(i,j)$ is called an edge if \hspace{2mm} $a_{ij}>0$.
    \item A neighbor of a vertex $i$ is another vertex $j$ such that $(i,j)$ is an edge.
    \item The degree of a node $i$ is the number of edges that connect the node $i$. i.e. Degree of $i=\big|\{j \hspace{2mm} | \hspace{1mm} i <j, \hspace{1mm} (i,j)\hspace{1mm} \text{is an edge}\}\big|$
    \item A path between two vertices $(i,j)$ is a sequence of pairs of vertices  $(i=k_0,k_1),(k_1,k_2),\ldots,(k_{n-1},k_n=j)$ such that each $(k_i,k_{i+1})$ is an edge.
    \item A path is called a cycle if the initial and final vertices coincide.
   \item  A graph $\mathcal{G}$ is said to be connected if there exists a path between any two distinct vertices. 
    \item A graph $\mathcal{G}$ is called a tree if it does not have cycles. A node in a tree is called a leaf if it has degree 1.
\end{itemize}
The following facts about trees are recalled as they are used in a forthcoming section.
\begin{thm}
\label{treet} Every tree has at least one leaf. The graph formed by removing the leaves of a connected tree (along with their corresponding edges) is again a connected tree or empty.
\end{thm}
  Another related matrix of equal importance for a graph is the Laplacian matrix $L$ that is defined for a symmetric adjacency matrix $A$ as $L:=O-A$ where $O$ is diagonal matrix whose diagonal entries are defined as $[O]_{ii}:=deg(i)$. The following theorem establishes the relation between the spectral properties of the Laplacian matrix and the connectivity of the graph. Note that the Laplacian matrix being symmetric, always has real eigen values and all of them can be shown to be non-negative (refer \cite{BulloNS}).

\begin{thm}
Let $A$ be a symmetric matrix with non-negative weights and $L=O-A$. The following hold:
\begin{itemize}
    \item The graph described by $A$ is connected if and only if the second smallest eigen value of $L$, denoted by $\lambda_2(L)$, is strictly positive.
    \item If the graph is connected and if $\textbf{1}:=\begin{bmatrix} 1 & 1 & \cdots & 1  \end{bmatrix}_{1\times n}^T$, then $Lx=0\Leftrightarrow x\in span\{ \textbf{1}\}$.
\end{itemize}
 
\end{thm}

Another important matrix of interest is the incidence matrix $B$. For this to be defined, two things need to be done. First, the edges of the graph have to be labelled arbitrarily, say $e=1,2,\cdots,|E|$ (the notation for the number of edges $|E|$ has vertical lines surrounding the letter E so that it may not be confused with the quantity energy $E$ that will be defined in a future section). Second, an arbitrary direction has to be assigned to each edge such that every edge $e=(i,j)$ should be thought of  as directed from $i$ to $j$ or from $j$ to $i$. The direction labelled to a given edge can be arbitrary but once assigned, has to be consistent. Note that if a graph is connected with $n$ nodes, it will at least have $n-1$ edges. i.e. $|E|\geq n-1$.

Then, the incidence matrix $B$ is a $n\times |E|$ matrix which is defined as 

\begin{align}
    B_{ie}&=+1, \hspace{3mm} \text{if node $i$ is source of the edge $e$} \nonumber\\
    B_{ie}&=-1, \hspace{3mm} \text{if node $i$ is the sink of the edge} \nonumber  \\
    B_{ie}&=\hspace{1mm}0, \hspace{3mm} \text{otherwise} .\nonumber
 \end{align}

Let $D$ be a $|E|\times|E|$ diagonal matrix whose $(e,e)^{th}$ entry is defined as
\begin{align}
    D_{ee}&=a_{ij}, \hspace{3mm} \text{$(i,j)$ are the nodes for the edge $e$}. \nonumber
\end{align}
Then, the incidence matrix is related to the Laplacian matrix as stated in the following theorem.
\begin{thm}
  \begin{align}
      L=BDB^T.
  \end{align}
\end{thm}
 Also, another Theorem related to the rank of $L$ and the rank of $B$ is stated below. The proofs of both the Theorems are standard and can be found in  \cite{BulloNS}.
\begin{thm}
  The following statements are equivalent
  \begin{itemize}
      \item the graph $\mathcal{G}$ is connected.
      \item $rank(L)=n-1$.
      \item $rank(B)=n-1$.
  \end{itemize}
\end{thm}

Note that the factorization of the Laplacian matrix as $L=BDB^T$ where $D$ is a diagonal matrix of positive diagonal entries renders the positive semi-definiteness of $L$ (and hence the non-negativity of its eigen values), very immediate and transparent.

\subsection{Geometric and Group Theoretic Preliminaries}
Denote by $M$, a smooth manifold and by $T_pM$ and $T_p^*M$, the tangent and co-tangent spaces at a point $p\in M$. If $M$ is equipped with a smoothly varying metric , denoted by $\mathbb{I}$, then $(M,\mathbb{I})$ is called a Riemannian Manifold. Denote by $\nabla$, the unique symmetric Levi-Civita connection induced on $M$ by the metric $\mathbb{I}$. For a smooth real-valued function $V$,  its differential at a point $p$  is denoted by $dV_p$ and its gradient, by $grad_p(V)$. Given a smooth map between two manifolds $F:M\rightarrow N$, its derivative map at a point $m\in M$ is denoted by $DF_m : T_mM \rightarrow T_{F(m)}N$. 

Let a manifold be equipped with a Lie group structure with the group operation $*$ , group inversion $()^{-1}$ and group identity $e$. A Lie group is called Abelian if the group operation is commutative. We also now assume henceforth unconditionally that all Lie groups in this work are assumed to be connected. It is customary to denote the Lie group by $G$ and an arbitrary point in it by $g$. The tangent space of a Lie group at its identity is defined to be the Lie algebra and is denoted by $\mathfrak{g}$. Let the dual of the Lie algebra be denoted by $\mathfrak{g}^*$. 

Denote by $ad_g$, the adjoint $ad_g:G\rightarrow G$ defined as $ad_g(h)=g*h*g^{-1}$.

Let $L_g$ denote the left translation map, denoted by $L_g: G \rightarrow G$ and defined as $L_g(h)=g*h$. Since $L_g$ is a diffeomorphism on $G$, this enables us to transport all tangent and cotangent vectors to $\mathfrak{g}$ and $\mathfrak{g}^*$. Similarly, the right translation map $R_g:G\rightarrow G$, defined as $R_g(h)=h*g$, also is a diffeomorphism of $G$ and likewise enables one to transport all tangent and co-tangent vectors to $\mathfrak{g}$ and $\mathfrak{g}^*$.

If $X\in T_gG$, then it can be left translated to $\mathfrak{g}$ , denoted by $g^{-1}*X$ which is defined as $g^{-1}*X := (DL_{g^{-1}})_g X$. Similarly, it can also be right translated to $\mathfrak{g}$ , denoted by $X*g^{-1}$ which is defined as $X*g^{-1} := (DR_{g^{-1}})_g X$. Hence, now we can uniquely identify every tangent vector in $\mathfrak{g}$, every cotangent vector in $\mathfrak{g}^*$ and any tensor in Cartesian products of $\mathfrak{g}$ and $\mathfrak{g}^*$ (through the left or right translation map). When there is an inner product on $\mathfrak{g}$, it is possible to uniquely induce a metric on $G$ by left translations. Such a metric is called left-invariant and it satisfies $\mathbb{I}_g(X,Y)=\mathbb{I}_{h*g}(h*X,h*Y)$ for all $g,h \in G$ and for all $X,Y \in T_gG$. It is also likewise possible to induce a metric on $G$ by right translations. Such a metric is called right-invariant and it satisfies $\mathbb{I}_g(X,Y)=\mathbb{I}_{g*h}(X*h,Y*h)$ for all $g,h \in G$ and for all $X,Y \in T_gG$. A metric that is both left and right invariant is called a bi-invariant metric.  Hereafter, it is assumed that the metric $\mathbb{I}$ on a Lie group  is bi-invariant. For details regarding any result  quoted so far, refer \cite{FFBullo} or \cite{Khalil}.

The following results give the necessary and sufficient condition for the existence of a bi-invariant metric on a Lie group $G$.  
\begin{thm} \cite{Milnor}
  \label{spivak} A Lie group $G$ admits a bi-invariant metric if and only if it is a Cartesian product of a compact Lie group and an Euclidean space . i.e. $G=G_C\times \mathbb{R}^p$ where $G_C$ is a compact Lie group and $\mathbb{R}^p$ is the Euclidean space, considered as a group under vector addition. 
\end{thm}

In a Lie group with a bi-invariant metric, the metric is ad-invariant as well, that is, invariant under adjoint operator. This property is called ad-invariance, which is
\begin{align}
    \mathbb{I}_g(X_g,Y_g)&=\mathbb{I}_{h*g*h^{-1}}(h*X_g*h^{-1}, h*Y_g*h^{-1})\nonumber \\ &=\mathbb{I}_{h*g*h^{-1}}(ad_h X_g, ad_h Y_g).
\end{align}

A nice compatibility between the geometry of the group and its algebra ensues if there exists a bi-invariant metric. Some of them are presented here. 
\begin{thm} \cite{Milnor}.
  If a  Lie group $G$ admits a bi-invariant metric, then 
  the integral curves of left/right invariant vector fields      %through $e$ 
     are precisely the  geodesics 
     %passing through $e$
     under the bi-invariant metric. In other words, the one-parameter subgroups and the geodesics through $e$ coincide. In other words, the Riemannian geodesic exponential coincides with the group exponential (loosely, the one parameter subgroups exp(tX) are the straight lines passing through $e$ with direction $X$).
      
\end{thm}

Relations between the Lie bracket operation and the covariant derivative also ensue. A couple of them are stated below.
\begin{thm} \cite{Milnor}  In a Lie group $G$ with a bi-invariant metric with its correspnding induced Levi-Civita connection $\nabla$, if $X,Y$ are two left invariant vector fields, then
  \begin{align}
      \nabla_X Y =\frac{1}{2} [X,Y]. \label{rf1}
  \end{align}
\end{thm}

\begin{thm} \cite{Milnor}
  \begin{align}
      \langle Z, [X,Y] \rangle = \langle X, [Y,Z]\rangle = -\langle X, [Z,Y]\rangle. \label{rf2}
  \end{align}
\end{thm}
 As a Corollary of both the theorems in Eqns \eqref{rf1}-\eqref{rf2}, we have
\begin{cor}
Given a vector $Y \in \mathfrak{g}$, The linear map $L_Y:\mathfrak{g}\rightarrow \mathfrak{g}$ defined as 
\begin{align}
    L_Y(X) := \nabla_X Y = \frac{1}{2}[X,Y].
\end{align}
is skew-symmetric. 
In other words, it satisfies for any $Z\in \mathfrak{g}$,
\begin{align}
    \langle Z, L_Y(X) \rangle = - \langle X, L_Y(Z)\rangle \label{skss} .
\end{align}
(or equivalently, the matrix representation of $L_Y$ in an orthonormal basis of $\mathfrak{g}$ is skew-symmetric).
\end{cor}
\textbf{Proof:}\\ 
\begin{align*}
    \langle Z, L_Y(X) \rangle &= \frac{1}{2}\langle Z, [X,Y]\rangle =-\frac{1}{2}\langle X, [Z,Y]\rangle\\
    \text{from  \eqref{rf2} }\\
    &=-\frac{1}{2}\langle X, \nabla_Z Y \rangle = -\frac{1}{2}\langle X, L_Y(Z)\rangle.
\end{align*}

\subsection{Morse Theoretic Preliminaries}

Let $F:M\rightarrow \mathbb{R}$ be a smooth scalar function on a manifold $M$. A point $c\in M$ is called a \textit{critical point} of $F$ if the differential at $c$ vanishes, that is $dF_c=0$. A critical point $c$ is called \textit{non-degenerate}, if the \textit{Hessian matrix} of $F$ in any local coordinate system is full rank (or equivalently, does not have zero as any of its eigen values) \cite{Milnor}. Note that as the Hessian matrix is symmetric for any scalar function, %(due to the equality of mixed partial derivatives),
its eigen values are all real. So, if non-zero, they must be either positive or negative.
\begin{defn}
A function $F$ is called a \textbf{Morse function} if all of its critical points are non-degenerate
\end{defn}
 The following are some important results about Morse functions. 
\begin{thm}\cite{Milnor}
  Let $F$ be a Morse function and $c\in M$ be a critical point in a manifold $M$ of dimension $m$ and assume $F(q)=0$ by adding an appropriate constant to $F$. Let $p$ eigen values of the Hessian matrix of $F$ at $c$ be positive and the remaining $m-p$, negative. Then, there exists a local coordinate system $(x_1,x_2,\ldots,x_m)$ around $c$ such that $c$ is in the origin and 
  $F$ in that coordinate system is $F(x_1,x_2,\ldots,x_m)=x_1^2+x_2^2+\cdots x_p^2 - x_{p+1}^2-x_{p+2}^2 - \cdots -x_{m}^2$ \label{mnf} 
\end{thm}
As a corollary of Theorem \ref{mnf}, we have that if $c$ is a local minimum, then $p=m$ and one can render $F$ as $F(x)=x_1^2+x_2^2+\cdots +x_m^2$ in some local coordinates around $p$ (since a quadratic function has origin as its minimum if and only if all its coefficients are positive). As an easy corollary of the above Theorem \ref{mnf}, we state another result that follows.
\begin{cor}\cite{Milnor}
The critical points of a  Morse function $F$ are isolated. In other words, if $c$ is a critical point, there exists a neighborhood of $c$
such that $c$ is the only critical point in that neighborhood (there can be no continuum of critical points for a Morse function). If $M$ is compact, this implies that the number of critical points are finite. \label{finite} \end{cor}

For control problems involving regulation to a point $p\in M$, one is interested in a special class of Morse functions, called Polar Morse functions.
\begin{defn}
A function $F$ is called \textbf{Polar Morse function} based at point $e$, if it satisfies the following properties:

\begin{itemize}
    \item $F$ is Morse.
    \item $F$ attains its global minimum at $e$ (whose value can be assumed as zero by translation).
    \item All the other critical points (except $e$) are not local minima (their Hessians have atleast one negative eigen value).
\end{itemize}
\end{defn}

The polar Morse function based at $e$ is ideal for the control problem of regulation to $e$. Since $e$ is the global minimum of $F$, going to $e$ is equivalent to minimizing $F$. Since $F$ has no other local minima, one can use the negative  gradient of $F$ and do gradient descent to steer a point to the global minimum $e$. It has been proven by Morse in \cite{Morse} that such polar Morse functions always exist on any smooth manifold. They are also proven to exist on any smooth compact manifolds with boundary in \cite{kod}.

Now we state another theorem related to the asymptotic behavior of gradient dynamical systems which will be very helpful in the forthcoming sections. 
\begin{thm} \cite{BulloNS}
Let $F$ be a smooth scalar function on a smooth Riemannian manifold $M$ and let $k_P,k_D>0$. Consider a first-order or second-order negative gradient flow given by
\begin{align}
    \dot{\gamma}&=-k_P grad(F), \label{ford}\\
    \nabla_{\dot{\gamma}}\dot{\gamma}&=-k_P grad(F) - k_D \dot{\gamma} \label{sord},
\end{align}
respectively. 
If all the sublevel sets of $F$ are compact (this is true trivially if $M$ itself is compact), then
\begin{itemize}
    \item any trajectory $\gamma(t)$ of  \eqref{ford} satisfies
    \begin{align*}
        \lim_{t\rightarrow \infty} \gamma(t) = c^*.
    \end{align*} 
    \item any trajectory $(\gamma(t),\dot{\gamma}(t))$ of \eqref{sord} satisifies
    \begin{align*}
          \lim_{t\rightarrow \infty} (\gamma(t),\dot{\gamma}(t)) = (c^*,0).
    \end{align*}
   where $c^*$ is a critical point of the function $F$, that is, $grad_{c^*} (F)=0$.
\end{itemize}

% any trajectory of both the dynamical systems asymptotically attain zero velocity and converge to a limit point $\gamma^*$ where $\gamma^*$ is a critical point of $F$. i.e. $grad(V)_{\gamma^*}=0$.
\end{thm}

\subsection{Review of the Centre Manifold Emergence Theorem}

In this section, we recollect the centre manifold emergence theorem that is crucial in the stability analyses that ensue in future sections. Let us first recapitulate the notion of the Jacobian of a vector field in a Riemannian manifold. 

If $W$ is a smooth vector field on a Riemannian manifold $G$, then the Jacobian of $W$ at a point $p\in G$ is defined as a linear map $J_p W:T_pG\rightarrow T_pG$ as follows:

\begin{align}
    J_p W (v) &:= (\nabla_v W)(p) ,  
\end{align}
where $  v \in T_pG$.

This induces a bilinear form $J:T_pG\times T_pG \rightarrow \mathbb{R}$ as follows:
\begin{align}
    J_p W (v_1,v_2):= \langle v_2, (\nabla_{v_1}W)_p\rangle.
\end{align}

With this, we state the center manifold emergence theorem.
\begin{thm} \cite{gh} (Center-manifold emergence theorem) 
  Let $W$ be a smooth vector field in a Riemannian manifold $G$ and let $g^*$ be any of its equilibrium point ($W(g^*)=0$). Let the Jacobian of $W$ at $g^*$ be such that their eigen values are either zero or with negative real parts. Let $p$ of the eigen values of $J_{g^*}W$ be zero and the remaining eigen values have strictly negative real part. Let $\nu$ be the maximum real part of the non-zero eigen values. Then, there exists a locally-unique $p$-dimensional invariant  submanifold passing through $g^*$ and tangential to $ker(J_pW)$ called the centre manifold $M_{cen}$ such that in a neighborhood $N_{cen}$ around $g^*$, any integral curve of the vector field $W$ starting in $N_{cen}$ exponentially converges to an invariant solution in $M_{cen}\cap N_{cen}$ with rate $\nu$.
 
\end{thm}
%The proof of this theorem is standard and can be found in textbooks like \cite{gh}. 

\section{ The notion of G-polar Morse Functions}

When the manifold is a Lie group and the objective is to achieve consensus, we want polar Morse functions to satisfy additional group theoretic properties. The importance of these properties will be evident in the coming sections. We name such class of functions on a Lie group $G$ as \textbf{$G-$Polar Morse}. 

\begin{defn}
\label{gpm}
A function $V$ on a Lie group is called \textbf{$G-$Polar Morse} if it satisfies the following properties:
\begin{itemize}

    \item $V$ is a polar Morse function based at identity $e$ of $G$
    \item $\forall g\in G$, $V(g)=V(g^{-1})$   (inversion symmetry)
    % \item $\forall g,h \in G$, $V(gh)=V(hg)$  (commutative invariance)
    \item The set of all critical points, denoted by $C:=\{g\in G \hspace{3mm} |\hspace{1mm} dV_g=0\}$ is a discrete subgroup of $G$
    \item The left transported and right transported gradient vectors in the Lie algebra $\mathfrak{g}$ coincide. That is,
    \begin{align}
        g^{-1}*grad_gV = grad_gV*g^{-1} 
    \end{align} (Note that this property is trivially satisfied for Abelian Lie groups like $\mathbb{R}^n$ and $\mathbb{S}^1$ as the left translation and right translation maps coincide). 
\end{itemize}
\end{defn}

In the next section, we will see why such a notion is important for consensus theory on Lie group. In Table \ref{tab1}, examples of such $G$-Polar Morse functions on commonly encountered Lie groups are listed. One can easily verify that the given functions in $\mathbb{R}^n$ and $\mathbb{S}^1$ satisfy all the required properties in Definition \ref{gpm}. As an illustrative example, we comment on the properties just for the $SO(3)$ case.
\begin{itemize}
    \item $V(R)=trace(I-R)$ is a polar Morse function. 
    \item $V(R)=trace(I-R)=trace(I-R^T)=V(R^T)=V(R^{-1})$ as for orthogonal matrices, $R^T=R^{-1}$.
    % \item $tr(I-R_1R_2)=tr(I-R_2R_1)$ as trace is invariant under the cyclic reordering of matrices
    \item The four critical points can be verified easily to form a subgroup of $SO(3)$.
    \item One can verify that the gradient of $V$ at $R$, when left and right translated, coincide and equals $\frac{1}{2det(\mathbb{I})}\mathbb{I}(R-R^T)\mathbb{I}$. 
\end{itemize}

It turns out that even for the higher dimensional $SO(n)$, $V(R)=trace(I-R)$ is a $G-$polar Morse function. And in the unitary group $U(n)$, the function $V(R)=trace(I-Re[R])$ (where $Re$ stands for the real part) is a $G$-polar Morse function as well. The issue of the guarantee of existence of such $G$-polar Morse functions on any arbitrary compact Lie group is discussed in the conclusion section. 
\begin{table}

    \begin{center}
    \begin{tabular}{|c|c|c|}
    \hline
       \textbf{Lie Group}  & \textbf{Morse Function}  & \textbf{Critical Points}\\
       \hline
        ($\mathbb{R}^N, +$) & $\frac{1}{2}||x||^2$ & $\{\textbf{0}\}$ \\
        \hline
        ($\mathbb{S}^1, +_{2\pi}$) & $1-\cos\theta$ & $\{\textbf{0},\pm \pi\}$\\
        \hline
        ($SO(3), .$) & $\mathrm{trace}(I-R)$ & $\{\textbf{I},\mathrm{diag}(-1,-1,1),$\\
         & & $\mathrm{diag}(1,-1,-1), $\\
         & & $\mathrm{diag}(-1,1,-1)\}$\\
        \hline
            \end{tabular}
    \label{tab1}
    \caption{%\centering
    Examples of $G$-Polar Morse functions}
    \end{center}
    \end{table}

 Let the gradient vector $grad$ of a smooth scalar function $V$ on $G$, when left translated to the Lie algebra, be denoted by $\mathfrak{grad}V$, be referred to as the \textbf{left Lie gradient}, which is $\mathfrak{grad}_gV(X):=grad_g V(g*X)$. %Similarly, let the gradient vector $grad$ of a smooth scalar function on $G$, when right translated to the Lie algebra, be referred to as the \textbf{right Lie gradient}, which is $\mathfrak{grad}_g(X):=grad_g (X*g)$.
 Then, some important properties of $G$-Polar Morse functions ensue.

\begin{lem}
\label{adjs}
If $V$ is a G-polar Morse function, then its left gradient at a point $g$ in the Lie algebra, is adjoint invariant under $ad_g$. Mathematically, $\mathfrak{grad}_g V = g*\mathfrak{grad}_g V*g^{-1}$.
\end{lem}

\textbf{Proof:}\\ 

The left translated gradient and the right translated gradient of a $G$-polar Morse function, coincide at any point $g$. So,
\begin{align}
    \mathfrak{grad}_gV &= g^{-1}*grad_g(V)  \label{1}\\ &\text{and} \nonumber \\\mathfrak{grad}_gV &=grad_g(V)*g^{-1}  \label{2}.
\end{align}
From  \eqref{1}, we get $grad_gV=g*\mathfrak{grad}_gV$ and substituting it in \eqref{2}, we have
\begin{align*}
    \mathfrak{grad}_g V = g * \mathfrak{grad}_g V * g^{-1},
\end{align*}
and hence the result ensues.

Let us recollect some basic results in Lie groups that will be frequently used in the forthcoming sections. 
\begin{lem}
Let $g(t)$ be a smooth curve on $G$ with its velocity  $\dot{g}(t)$. Then, $\frac{d}{dt}g^{-1}(t) := \dot{g}^{-1}(t)=-g^{-1}*\dot{g}(t)*g^{-1}$.
\end{lem}
\begin{lem} 
\begin{align}
    \frac{d}{dt}(g_i*g_j)=g_i*\dot{g}_j + \dot{g}_i*g_j
    \end{align}
\end{lem}
% \textbf{Proof:}\\

% Since $g(t)*g^{-1}*(t)=e$ and constant, differentiating both sides and applying chain rule, we get the required result after rearranging.
\vspace{2mm}
\begin{thm}
For a $G-$Polar Morse function, we have 
\begin{align}
    g^{-1} * grad(V)_g = - g * grad(V)_{g^{-1}} \label{graf}
\end{align}
or equivalently
\begin{align}
    \mathfrak{grad}_gV = -\mathfrak{grad}_{g^{-1}}V \label{11}.
\end{align}
\end{thm}

The physical significance of \eqref{11} is that the gradients at $g$ and $g^{-1}$ are negatives of each other, when viewed in the Lie algebra.

\textbf{Proof:}\\ 

Consider an arbitrary tangent vector at $g$ that is the velocity of a curve, that is denoted by $\dot{g}$. From the property of inversion symmetry, we have
\begin{align}
    V(g)&=V(g^{-1})
.\end{align}
Taking the time derivative of $V$ along the curve $g(t)$ and using chain rule, we have
\begin{align}
     \langle grad(V)_g ,\dot{g}\rangle &= \langle grad(V)_{g^{-1}} ,\dot{g^{-1}}\rangle,\\
    \Longrightarrow \langle grad(V)_g , \dot{g}\rangle &= \langle grad(V)_{g^{-1}} ,-g^{-1}*\dot{g}*g^{-1}\rangle.  \label{prevv} 
\end{align}
By left invariance of the metric, translating all vectors to the Lie algebra through left translation by $g$ on LHS of \eqref{prevv} and by $g^{-1}$ on RHS of \eqref{prevv}, we get
\begin{align}
    \langle g^{-1}*grad(V)_g , g^{-1}*\dot{g} \rangle = - \langle g * grad(V)_{g^{-1}} , \dot{g}*g^{-1} \rangle
\end{align}
But 
\begin{align*}
    &\dot{g}*g^{-1}=(g*g^{-1})*\dot{g}*g^{-1}=\\&g*(g^{-1}*\dot{g})*g^{-1}= ad_g (g^{-1}\dot{g}).
\end{align*}
Denoting $g^{-1}*\dot{g}\in \mathfrak{g}$ by $X$, we have
\begin{align}
  \langle g^{-1}*grad(V)_g , X \rangle = - \langle g * grad(V)_{g^{-1}} ,ad_g X \rangle.
\end{align}
Hence,
\begin{align}
  \langle \mathfrak{grad}_g V , X \rangle = - \langle \mathfrak{grad}_{g^{-1}} V ,ad_g X \rangle \label{just}.
\end{align}
Taking $ad_{g^{-1}}$ for both vectors in the RHS of \eqref{just}, and using the ad-invariance (Lemma \ref{adjs}), we have,
\begin{align}
    \langle \mathfrak{grad}_g V , X \rangle &= - \langle ad_{g^{-1}} \mathfrak{grad}_{g^{-1}} V , \overbrace{ad_{g^{-1}}ad_g}^\text{=identity}X   \rangle \nonumber \\
    &=- \langle ad_{g^{-1}} \mathfrak{grad}_{g^{-1}} V , X   \rangle \nonumber\\
    &=-\langle  \mathfrak{grad}_{g^{-1}} V , X   \rangle,
\end{align}
which yields the requires result.

\begin{figure}[h]
    \centering
    \includegraphics[scale=0.4]{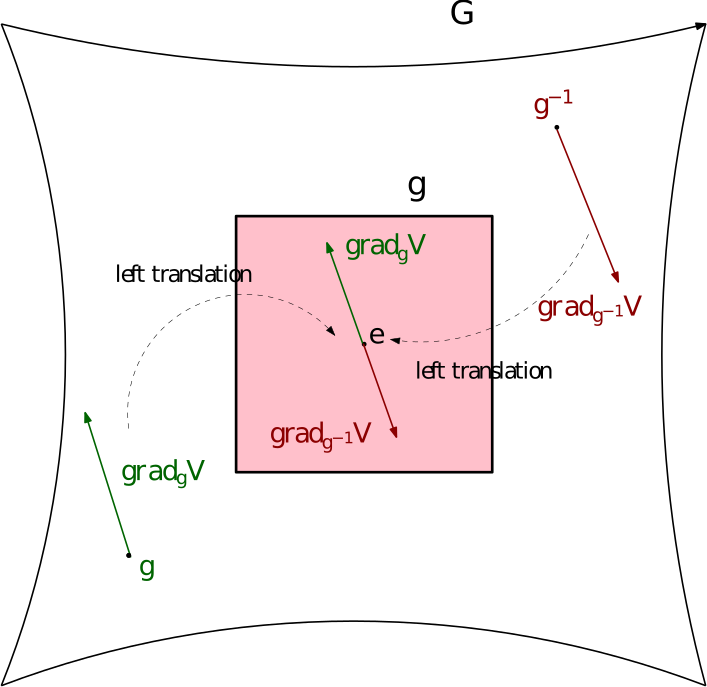}
    \caption{Illustration of $\mathfrak{grad}_g V = -\mathfrak{grad}_{g^{-1}} V$}
    \label{fig:fig1}
\end{figure}

 The result is visualized in Figure  \ref{fig:fig1}.

\section{Group Theoretic Insights:\\ Review of Laplacian Flow and Coupled Oscillators}

In this section, the Laplacian flow and coupled oscillators are reviewed and are cast in the $G-$polar Morse theoretic framework, proposed in Section 3. This enables one to generalize the theory of consensus and synchronization to general Lie groups admitting a bi-invariant metric and a $G$-polar Morse function. 

\subsection{Laplacian Flow}
The Laplacian flow is a standard algorithm for consensus in Euclidean space. Denoting by $x_i$, the configuration of the $i^{th}$ agent in a Euclidean space $\mathbb{R}^N$, we have the following Laplacian flows for consensus

\begin{align}
    \text{First-Order Laplacian Flow:} \nonumber
    \\
    \dot{x}_i=- k_P\sum_{j=1}^n a_{ij}(x_i-x_j) \label{folf}\\
      \text{Second-Order Laplacian Flow:} \nonumber \\
          \ddot{x}_i=-k_P\sum_{j=1}^n a_{ij}(x_i-x_j)-k_D \dot{x}_i \label{solf}
\end{align}
where $k_P,k_D>0$.

The crucial component of these algorithms is the inter-agent consensus term $(x_i-x_j)$. It steers the agent $i$ towards agent $j$. 

The important observation one notes here is that by considering the $G$-polar Morse function on Euclidean space as $V=\frac{1}{2}||x||^2$ , the consensus term can be written as 
\begin{align}
    (x_i-x_j)&=grad_{i} V(x_i-x_j) .
    %\\
    %V&=\frac{1}{2}||x||^2. \nonumber
\end{align}
 This is natural because the negative gradient of a polar Morse function like $V$ drives the flow towards the minimum of $V$, which happens when $x_i-x_j=0$ or  $x_i=x_j$.

Both the first and second-order Laplacian flow lead to consensus from all initial conditions if the graph is connected and the gains $k_P,k_D$ are strictly positive \cite{BulloNS}.

 The general intuition as to why this is true can be explained by a mechanical interpretation. Each edge with an interconnection $a_{ij}>0$ can be viewed as a spring connecting agents $i$ and $j$ with a spring constant $a_{ij}$ and with the standard quadratic potential energy $a_{ij} V(x_i-x_j)=a_{ij} \frac{1}{2}||x_i-x_j||^2$. The total potential energy of the system is given by $V_{TOT}=\frac{1}{2}\sum_{i,j} a_{ij}V(x_i-x_j)$ (since each spring is counted twice, there is a half in the front). Since the network is connected, the system arrives at a consensus due to the effect of the spring, and the damping term $\gamma_D \dot{x}_i$ for second-order agents.
 
The entire dynamics in \eqref{folf}-\eqref{solf} can be rewritten as a gradient dynamical flow as

\begin{align}
    \dot{x}_i &= -k_P grad_i (V_{TOT}),\\
    \ddot{x}_i &=-k_P grad_i(V_{TOT}) - k_D \dot{x}_i.
\end{align}

 Formally, it has been shown  that
 
 \begin{thm} \cite{BulloNS}
 Subject to the following assumptions
 \begin{itemize}
     \item $k_P,k_D>0$,
     \item the graph describing the agents are undirected and connected,
     \item the adjacency matrix is symmetric,
 \end{itemize}
  the first/second-order Laplacian flow starting from any initial condition, satisfies for all $i,j$,
     \begin{align*}
      \lim_{t\rightarrow \infty} \dot{x}_i(t)&=0, \hspace{2mm}
       \\\lim_{t\rightarrow \infty}x_i(t) & \hspace{2mm}  \text{exists}   ,    
     \\         \lim_{t\rightarrow \infty} x_i(t)&=\lim_{t\rightarrow \infty} x_j(t) \hspace{3mm}  \hspace{2mm} .
              \end{align*}
 \end{thm}

\subsection{Coupled Oscillators: Consensus on the circle}

We now consider the standard model of coupled identical oscillators with zero natural frequency. Let $\theta\in \mathbb{R}$ be the standard angular coordinate of the unit circle. 
\begin{align}
    \text{First-Order Coupled Oscillators:} \nonumber
    \\
    \dot{\theta}_i=- k_P\sum_{j=1}^na_{ij}\sin(\theta_i-\theta_j) \label{sfo}\\
      \text{Second-Order Coupled Oscillators:} \nonumber \\
          \ddot{\theta}_i=-k_P\sum_{j=1}^na_{ij}\sin(\theta_i-\theta_j)-k_D \dot{\theta}_i \label{sso}
\end{align}
where $k_P,k_D>0$.

For the first-order case and when $a_{ij}=\frac{1}{n}$ for all  $i,j$, we have the Kuramoto oscillator
\begin{align}
     \text{Kuramoto Oscillator} \nonumber
    \\
    \dot{\theta}_i=- k_P \sum_{j=1}^n\frac{1}{n} \sin(\theta_i-\theta_j).
\end{align}

Once again the crucial term for consensus is the inter-agent sinusoidal term $\sin(\theta_i-\theta_j)$. As mentioned before,  this is also the gradient of a $G$-polar Morse function on the circle. Considering $V=1-\cos\theta$, we get

\begin{align}
    \sin(\theta_i-\theta_j)&=grad_{i} V(\theta_i-\theta_j) \\
    V&=1-\cos\theta \nonumber.
\end{align}

The same mechanical intuition can be given with $a_{ij}(1-\cos(\theta_i-\theta_j))$ being interpreted as the potential energy of a spring connecting agents $i$ and $j$ and once again the entire system can be cast in gradient form by defining $V_{TOT}=\frac{1}{2}\sum_{i,j}V(\theta_i-\theta_j)$. The systems now become

\begin{align}
    \dot{\theta}_i &= -k_P grad_i (V_{TOT}),\\
    \ddot{\theta}_i &=-k_P grad_i(V_{TOT}) - k_D \dot{\theta}_i.
\end{align}

So we see that both the oscillator dynamics for consensus on $\mathbb{S}^1$ and Laplacian flow for consensus on Euclidean space have the same geometric and group theoretic structure. Here too, the Morse function mimics the potential energy of a spring (it is minimum at identity, and does not have a local minimum anywhere else).  

Note that the crucial difference from the Euclidean case is that while the Morse function there had no other critical points apart from the global minimum at identity, the Morse function on the circle $1-\cos\theta$ has another critical point which is $\pi$. The set of critical points $\{0,\pi\}$ are a two-element subgroup. This cannot be avoided as $\theta=\pi$ is the maximum of $V$ and any continuous function on a compact set like the circle $\mathbb{S}^1$ has to have a maximum (this shows that for compact manifolds, one cannot escape having critical points other than the global minimum). 
\subsubsection{Comment on the subgroup property of the critical points}

We know that in a gradient descent flow, it is possible to converge to a critical point other than the global minimum.
% but these other critical points are unstable if their Hessian atleast has one negative value (a saddle or a maxima). So, unless the initial condition lies in the stable manifolds of such saddles or maxima (which are measure zero), for almost all initial conditions (the word 'almost' is used in a measure theoretic sense), the trajectories converge to the minimum. But it is possible to converge to the other critical points as well (for initial conditions of zero measure) and 
So, in general, at equilibrium, in addition to the consensus solution $\theta_i=\theta_j$ for all $i,j$, we can also have 
\begin{align}
    \theta_i-\theta_j \in C=\{0,\pi\} \label{c}.
\end{align}

So, in limit, it is possible for the oscillators to be either in consensus $\theta_i=\theta_j$ or anti-consensus $\theta_i-\theta_j=\pi$ with one another. 
But it is important that $C$ be a subgroup for \eqref{c} to be consistent. Because,

\begin{align}
    \overbrace{\theta_i-\theta_j}^\text{$\in C$}=\overbrace{(\theta_i-\theta_k)}^\text{$\in C$}+\overbrace{(\theta_k-\theta_j)}^\text{$\in C$} \label{sas}.
\end{align}

Only if $C$ is a subgroup, the above equation \eqref{sas} is consistent as the right hand side is the sum of two elements in $C$ and it has to be remain in $C$ as the LHS is in $C$. So, $C$ has to be closed under $+$ which is the group operation in circle. This is required for a transitive closure of the consensus/anti-consensus relation which is: if agent $i$ is in consensus/anti-consensus with agent $j$ and if agent $j$ is in consensus/anti-consensus with agent $k$, then agents $i$ and $k$ are also in consensus/anti-consensus with each other.

\subsubsection{Comments on the graph structure}

We expect the global minimum (where all the oscillators are in consensus with each other) to be locally exponentially stable, the anti-consensus solutions (where some of the oscillators are anti-consensus with the others) to be unstable, and we would require that there are no other equilibria of the system (critical points of $V_{TOT}$).  This would render the globally minimal synchronous solution almost globally asymptotically stable. But while the first two expectations are fulfilled for arbitrary graphs, the third expectation that there are no other equilibria other than the consensus/anti-consensus equilibria is not the case for arbitrary graphs. So, the next following question would be that are the other equilibria that are neither conesus nor anti-consensus, locally unstable? This is also not true for all graphs and the cyclic graph with 5 nodes is proven to exhibit an equilibrium that is locally attracting. For what kinds of graphs, is the consensus equilibrium, also the only attracting equilibrium (that renders it almost globally asymptotically stable)? We shall now review the convergence properties of  \eqref{sfo}-\eqref{sso} under different graph structures. 

First let us define some notions related to the geometry of the circle.
\begin{defn}
Let $0\leq \gamma< \pi$
\begin{itemize}
    \item The arc subset $\bar{\Gamma}_{arc}(\gamma)$ is defined as the set of agent angles $\theta_i$ such that there exists an arc of length $\gamma$ in $\mathbb{S}^1$ containing all the angles $\theta_i$. The set $\Gamma_{arc}(\gamma)$ is defined to be the interior of the set $\bar{\Gamma}_{arc}(\gamma)$.
    \item Given a graph $\mathcal{G}$, a cohesive subset $\Delta^{\mathcal{G}}(\gamma)$ is defined as the set of agent angles $\theta_i$ such that $\theta_i,\theta_j$ are contained in an arc of length $\gamma$ if $(i,j)$ is an edge of the graph 
\end{itemize}
\end{defn}

We now capture the various notions of synchronization through formal definitions below.
\begin{defn}
Angles  $\theta_i(t)$ evolving on $\mathbb{S}^1$ are said to be
\begin{itemize}
 \item \textbf{phase synchronized} if  $\theta_i(t)=\theta_j(t)$ $\forall \hspace{1mm} t$  , $\forall \hspace{1mm} i,j$\item  \textbf{frequency synchronized} if  $\dot{\theta}_i(t)=\dot{\theta}_j(t)$ $\forall t$, $\forall\hspace{1mm} i,j$
 \item \textbf{phase cohesive} if  $\forall\hspace{1mm} t$, either $\theta_i(t)\in \Gamma_{arc}(\gamma)$ or with respect to a graph $\mathcal{G}$, $\theta_i(t)\in \Delta^{\mathcal{G}}(\gamma)$
\end{itemize}
\end{defn}

\begin{thm} \cite{DORFLER20141539}
\label{Dorf1} Consider the dynamic equations \eqref{sfo} or \eqref{sso} with a connected graph with symmetric  adjacency matrices and $k_P,k_D>0$. Then,
\begin{itemize}
    \item For all initial conditions $\theta_i(0)$, the angles converge to the set of critical points of $V_{TOT}$. 
    \item If the initial angles are phase cohesive w.r.t some $0\leq \gamma< \pi$, then the solution remains phase cohesive.
    \item The consensus solution is locally exponentially stable with an exponent $-\lambda_2(L)\cos(\frac{\gamma}{2})$ for the domain-of-attraction being chosen as $\Gamma_{arc}(\gamma)$ for any $0\leq \gamma < \pi$.
    \item The consensus solution is almost globally asymptotically stable (it is the only attracting equilibrium) for the following graphs \begin{enumerate}
        \item a tree   \cite{inbook}
        \item a sufficiently dense graph \cite{soa2012},\cite{soa2020}, \cite{soa2020first} (in particular, this includes the complete graph \cite{BulloNS},  \cite{DORFLER20141539})  - the current state of the art is that for a graph of $n$ nodes, if the degree of each node $d_i$ satisfies $d_i>0.7889n$ for all $i$, then the consensus equilibrium is the only attracting equilibrium 
       
    \end{enumerate}
\end{itemize}
\end{thm}

It is also known that there are graphs for which almost global asymptotic stability of the consensus does not hold \cite{ce} where there are other attracting equilibria (or non-consensus local minima of $V_{TOT}$).

% Formally, it can be shown that

%  \begin{thm}
 
%   If $k_P,k_D>0$, the graph describing the agents are connected and the adjacency matrix is stochastic and symmetric, then the first/second-order dynamics of coupled oscillators, starting from any initial condition, satisfies asymptotically and for all $i,j$,
%      \begin{align}
%       lim_{t\rightarrow \infty} \dot{\theta}_i(t)&=0 
%      \\lim_{t\rightarrow \infty}\theta_i(t) & \hspace{2mm}  \text{exists}       
%      \\
%      (lim_{t\rightarrow \infty} \theta_i(t)-lim_{t\rightarrow \infty} \theta_j(t)) \in C&=\{0,\pi\}
%               \end{align}
%  \end{thm}

\subsection{Synchronization of Coupled Oscillators}

For synchronization of oscillators, we consider the case wherein each oscillator also has its own natural frequency $\omega_i$ apart from having to arrive at a consensus with other oscillators. In this case, the consensus dynamics in \eqref{sfo} gets modified by the addition of an extra natural frequency which becomes
\begin{align}
    \label{syncos}
    \dot{\theta}_i= \omega_i -k_P \sum_{j=1}^na_{ij}\sin(\theta_i-\theta_j),
\end{align}

which again can be recast as 
\begin{align}
     \label{syncoss}
    \dot{\theta}_i= \omega_i -k_P\sum_{j=1}^na_{ij} grad_i V(\theta_i-\theta_j).
\end{align}

Here each oscillator has two driving velocities: $\omega_i$ that makes it oscillate with its own preset constant angular frequency and another consensus term that drives it towards matching the state of its neighboring oscillators. The equilibria of such systems as well have been studied in detail \cite{BulloNS}. Let $\omega_m=\frac{1}{n}\sum_i \omega_i$ (mean natural frequency). Then, subtracting $\omega_m$ both sides and defining $\Delta\omega_i=\omega_i-\omega_m$, we get
\begin{align}
    \dot{\theta}_i-\omega_m=\Delta\omega_i-\sum_{j=1}^na_{ij}\sin(\theta_i-\theta_j) \label{d}.
\end{align}

Let us situate ourselves in a frame that is rotating with the mean angular velocity $\omega_m$. Then in this frame, the oscillators evolve by the RHS of \eqref{d}.

It is intuitvely expected that if the coupling strength is much larger than $\Delta \omega_i$, the oscillators should get in sync eventually and if the coupling strength is much smaller than $\Delta\omega_i$, the oscillators should not get in sync. Formally one can show the following result:
\begin{thm} \cite{DORFLER20141539} Define
\begin{align}
   ||\Delta\omega||_2:=\sqrt{\frac{1}{2}\sum_{i,j=1}^n (\Delta\omega_i-\Delta\omega_j)^2}, \end{align}
   and $\lambda_2(L)$ as the second smallest eigen value of the symmetric matrix $L=O_{n\times n}-A$. If $k_P \lambda_2(L)>||\Delta\omega||_2$, then frequency synchronization is achieved for all oscillators asymptotically to the mean frequency $\omega_m$ if their initial spread is restricted to an arc (around the synchronous equilibrium), whose length is sufficiently small. Phase synchronization is achieved only when all the natural frequencies are equal.        \label{Dorf}
\end{thm}
%The proof can be found in \cite{DORFLER20141539}.

\section{Consensus on Lie groups }
\subsection{The protocol}
We now consider a  consensus protocol on a Lie group that generalizes the Laplacian flow in Euclidean space and the  sinusoidally coupled oscillators on the unit circle. We assume the Lie group to be compact hereafter. This is not a problem as already it was seen that any Lie group with a bi-invariant metric can be written as a Cartesian product of a compact Lie group and an Euclidean space. We can take care to design consensus and synchronization on Euclidean space separately and concentrate only on the compact Lie group. In a Lie group, all velocities can be left translated to the Lie algebra $\mathfrak{g}$ as visualized below. Henceforth, it is assumed that all tangent vectors are shown after left transport to the Lie algebra, including the velocity of each agent. 

\begin{figure}[h]
	\center{\includegraphics[scale=0.2]{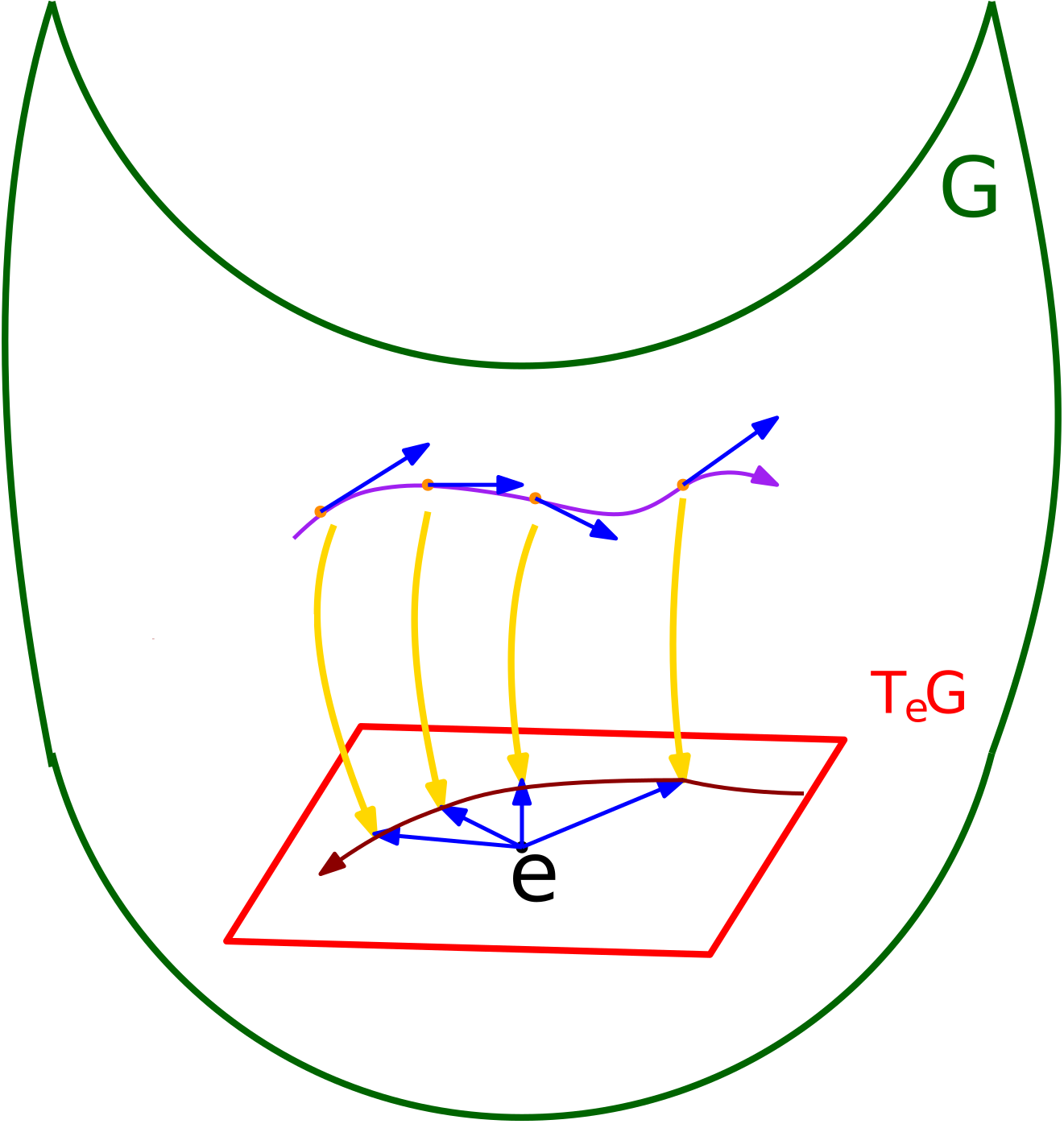}}
	\caption{Visualizing velocity of a curve in the Lie Algebra}
	\label{fig:figg3}
\end{figure}

As already defined, the left invariant error of agent $i$ w.r.t agent $j$ as $E_{ij}=g_j^{-1}*g_i$. Note that this inter-agent error is identity when the agents are coincident and deviate from identity otherwise. This is visualized below.
\begin{figure}[h]
	\center{\includegraphics[scale=0.2]{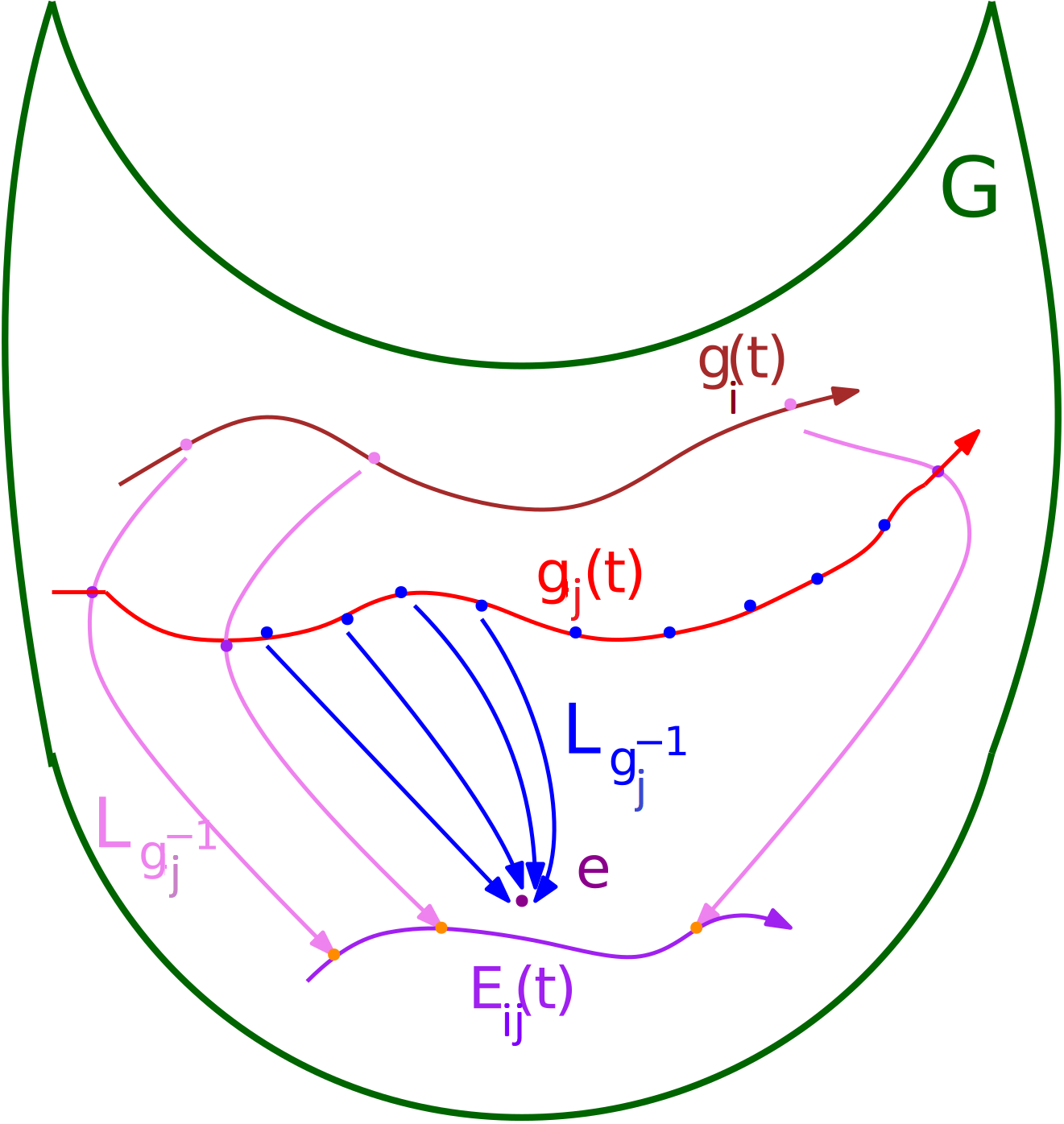}}
	\caption{Visualizing $E_{ij}$}
	\label{fig:figg2}
    \end{figure}

Next, we need to capture the deviation of this error from identity. That is what is exactly done by a $G$-polar Morse function. So, $V(E_{ij})$ will be zero if and only if $g_i=g_j$ and strictly positive otherwise. So, we use $V(E_{ij})$ to construct the consensus dynamics.  

We first prove various elementary properties of $V_{ij}: G\times G \rightarrow \mathbb{R}$ defined by $V_{ij}:=V(E_{ij})=V(g_j^{-1}*g_i)$. Recall the notation that the gradient translated to the Lie algebra be denoted by $\mathfrak{grad}$ - that is, $\mathfrak{grad}_g V (v)=grad_g V (g*v)$. Let $\mathfrak{grad}^i V_{ij}$ and $\mathfrak{grad}^j V_{ij}$ denote the gradients of $V_{ij}$ with respect to the variables $g_i$ and $g_j$ respectively. Then, we have

\begin{thm} 
\label{oppt}

\begin{align}
\label{opp}
    \mathfrak{grad}^i V_{ij} = -\mathfrak{grad}^j V_{ji}.
\end{align}
%where the superscript $i$ or $j$ indicates w.r.t what variable, gradient is being taken - i.e. $g_i$ or $g_j$.
\end{thm}

\textbf{Proof:}\\

Let $\dot{g}_i(t),\hspace{1mm}\dot{g}_j(t)$ be the velocities of curves. Then, by chain rule, we have
\begin{align*}
    &\frac{d}{dt}V_{ij} \nonumber 
    \\&=\langle grad_{g_j^{-1}*g_i} V,  \frac{d}{dt}(g_j^{-1}*g_i)\rangle\\
    &=\langle grad_{g_j^{-1}*g_i} V, \dot{g}_j^{-1}*g_i+g_{j}^{-1}*\dot{g}_i\rangle\\
    &=\langle grad_{g_j^{-1}*g_i} V  ,-g_j^{-1}*\dot{g}_j*g_j^{-1}*g_i+g_j^{-1}*\dot{g}_i\rangle.
    \end{align*}
    When we left multiply both the gradient and the velocities by $(g_j^{-1}*g_i)^{-1}=g_i^{-1}*g_j$, they both will be in the Lie algebra. Doing this and simplifying, we have
    \begin{align}
    &=\langle\overbrace{ \mathfrak{grad}_{g_j^{-1}*g_i} V , -g_i^{-1}*\dot{g}_j*g_j^{-1}*g_i + g_i^{-1}*\dot{g}_i}^\text{left transported to $\mathfrak{g}$}\rangle.\end{align}
    Using adjoint operator, the velocities simplify further as
    \begin{align}
    &=\langle \mathfrak{grad}_{g_j^{-1}*g_i} V ,-ad_{g_i^{-1}*g_j}\underbrace{g_j^{-1}*\dot{g}_j}_\text{$\in \mathfrak{g}$} + \underbrace{g_i^{-1}*\dot{g}_i}_\text{$\in \mathfrak{g}$}\rangle. \label{cccc}
    \end{align}
    Now, since gradient is linear, we can split the sum into two parts as 
    \begin{align}
        =&\langle \mathfrak{grad}_{g_j^{-1}*g_i} V ,-ad_{g_i^{-1}*g_j}{g_j^{-1}*\dot{g}_j} \rangle + \nonumber\\ &\langle \mathfrak{grad}_{g_j^{-1}*g_i} V , g_i^{-1}*\dot{g}_i \rangle.
    \end{align}
    
     Since the metric is ad-invariant, adjointing the first-term by $ad_{g_j^{-1}*g_i}$, we get
     \begin{align}
     &= \langle ad_{g_j^{-1}*g_i} \mathfrak{grad}_{g_j^{-1}*g_i} V, - \overbrace{ad_{g_j^{-1}*g_i} ad_{g_i^{-1}*g_j}}^\text{yields identity} g_j^{-1}*\dot{g}_j \rangle \nonumber\\
     &=\langle ad_{g_j^{-1}*g_i} \mathfrak{grad}_{g_j^{-1}*g_i} V, -  g_j^{-1}*\dot{g}_j \rangle \label{aaa}.
     \end{align}
But by adjoint invariance of the gradient of a $G-$polar Morse function in Lemma \ref{adjs}, we have $ad_{E_{ij}} \mathfrak{grad}_{E_{ij}} V =\mathfrak{grad}_{E_{ij}} V $. Substituting this in the above equation \eqref{aaa}, we get
\begin{align}
    &\langle ad_{g_j^{-1}*g_i} \mathfrak{grad}_{g_j^{-1}*g_i} V, -  g_j^{-1}*\dot{g}_j \rangle = \langle  \mathfrak{grad}_{g_j^{-1}*g_i} V, -  g_j^{-1}*\dot{g}_j \rangle \label{bbb}.
\end{align}
Hence, ignoring the adjoint term in \eqref{cccc} due to \eqref{bbb}, we get
\begin{align}
    \frac{d}{dt}V_{ij} &= \langle \mathfrak{grad}_{g_j^{-1}*g_i} V ,-g_j^{-1}*\dot{g}_j + g_i^{-1}*\dot{g}_i\rangle.
\end{align}
Let $v_i=g_i^{-1}*\dot{g}_i$ , $v_j=g_j^{-1}*\dot{g}_j$ be the velocity of the $i^{th}$ and $j^{th}$ agents left translated to Lie algebra. Then we have
\begin{align}
    \frac{d}{dt}V_{ij} = \mathfrak{grad}_{g_j^{-1}*g_i} V (v_i-v_j) \label{hl}.
\end{align}

Since we want the gradient of $V_{ij}$ with respect to just the $i^{th}$ agent, we set $v_j=0$ in \eqref{hl} and hence we have
\begin{align}
    \frac{d}{dt}V_{ij}\bigg|_{v_j=0} &= \mathfrak{grad}_{g_j^{-1}*g_i} V (v_i) \\ \Rightarrow  \mathfrak{grad}^i V_{ij} &= \mathfrak{grad}_{g_j^{-1}*g_i} V.
\end{align}

Interchanging the indices $i,j$, we have
\begin{align}
    \mathfrak{grad}^j V_{ji} = \mathfrak{grad}_{g_i^{-1}*g_j} V.
\end{align}

Due to inversion symmetry \eqref{graf},  in the Lie algebra, the gradients at inversely related points are negatives of each other. But $g_i^{-1}*g_j$ and $g_j^{-1}*g_i$ are inverses of each other. So,
\begin{align}
    \mathfrak{grad}^j V_{ji} = \mathfrak{grad}_{g_i^{-1}*g_j} V = - \mathfrak{grad}_{g_j^{-1}*g_i} V = - \mathfrak{grad}^i V_{ij},
\end{align}
which yields the required result.  

We are now ready to propose the consensus algorithm on a Lie group, for both first- and second-order agents. With $v_i$ being the velocity of the $i^{th}$ agent in the Lie algebra and $\nabla$ being the bi-invariant connection induced by the bi-invariant metric $\mathbb{I}$, we have
\begin{align}
    \text{First-Order Laplacian Flow:} \nonumber\\
    v_i = -k_P \sum_{j=1}^n a_{ij} \hspace{1mm} \mathfrak{grad}^i_{E_{ij}} V_{ij} ,\label{gfolf} \\
    \text{Second-Order Laplacian Flow:} \nonumber\\
    \nabla_{v_i}v_i = -k_P \sum_{j=1}^n a_{ij} \hspace{1mm} \mathfrak{grad}^i_{E_{ij}} V_{ij} - k_D  v_i \label{gsolf}.
\end{align}
Recall that $\nabla_{v_i}v_i$ is the acceleration of the $i^{th}$ agent in Riemannian geometry and hence the above equations are direct generalizations of the Laplacian flow and sinusoidal consensus protocols to an arbitrary Lie group.

\subsection{Equilibrium Analysis}

For the first-order system in \eqref{gsolf}, the equilibriua are the configurations that satisfy
\begin{align}
   \forall \hspace{1mm}i: \hspace{2mm} \sum_{j=1}^na_{ij} \hspace{1mm} \mathfrak{grad}^i_{E_{ij}} V_{ij} = 0 \label{eqm}.
\end{align}

Included within these equilibria are:
\begin{itemize}
    \item \textbf{Consensus Solution:}        Where $g_i=g_j$ for all $i,j$ which implies $\mathfrak{grad}^i_{E_{ij}} V_{ij}=0$ individually for every $i,j$.
    
    \item \textbf{Anti-Consensus Solutions:} 
    Where $g_i^{-1}*g_j\in C$ for all $i,j$ which also implies $\mathfrak{grad}^i_{E_{ij}} V_{ij}=0$ individually for every $i,j$.
    
    \item \textbf{Other Solutions:} 
    Where the  gradients $\mathfrak{grad}^i_{E_{ij}} V_{ij}=0$  do not individually vanish but collectively satisfy the equilibrium condition \eqref{eqm}. 
    
\end{itemize}
For the sake of ease of reference, let us call the equilibria where the gradients individually vanish (consensus and anti-consensus points) as \textbf{trivial equilibria} and the other equilibria as \textbf{non-trivial equilibria}.

For the second-order system in \eqref{gsolf}, the dynamics in state-space form is
\begin{align}
    \dot{g}_i &= g_i*v_i, \nonumber \\
    \nabla_{v_i}v_i &= -k_P \sum_{j=1}^n a_{ij} \hspace{1mm} \mathfrak{grad}^i_{E_{ij}} V_{ij} - k_D  v_i.
\end{align}

At equilibrium, we hence have $v_i=0$ and the same condition \eqref{eqm} for the first-order case. So, the equilibria of the second-order flow are those where the velocities of the agents vanish and the positions same as that of the first-order flow. 

Ideally, we would only want the consensus solution to exist and be globally asymptotic stable. But as we have seen, for compact manifolds, $C$ 
%(set of critical points of a Morse function)
has points other than identity and hence we have anti-consensus solutions as well, similar to oscillators.

It can be shown (it will be shown in a future section) that the consensus points are locally exponentially stable and that these anti-consensus points possess an unstable direction and hence cannot be locally stable (these are done in the next section). So, then, if we do not have any other non-trivial equilibria, then the locally exponentially stable consensus equilibrium can be almost globally exponentially stable. But as is known from the theory of oscillators, this is not the case for all graphs. There can be other non-trivial solutions to \eqref{eqm} where the individual gradients in the summation do not vanish.

As it was mentioned, in the theory of oscillators, it is known that for the Kuramoto oscillator where the graph is a complete graph with equal weights ($a_{ij}=\frac{1}{n}$ for all $i,j$), there are no other non-trivial equilibrium solutions and that the consensus equilibrium is almost globally asymptotically stable . This has also been established for trees. Now, we will establish that this is the case here for general Lie groups as well, when the graph is a tree. We will start the analysis with simpler graphs like two-agent system and a line graph as precursors to visualize the situation geometrically. 
\subsubsection{Two Agent case}
Consider the case when there are only two agents. Then, the consensus protocols generalize as follows:

\begin{itemize}
    \item First-Order:
    \begin{align}
        v_1 = -\frac{k_P}{2}\mathfrak{grad}^1 V_{12} .\\
        v_2= - \frac{k_P}{2}\mathfrak{grad}^2 V_{21}.
    \end{align}
    \item Second-Order:
    \begin{align}
        \nabla_{v_1}v_1 = -\frac{k_P}{2}\mathfrak{grad}^1 V_{12} - k_D v_1,\\
        \nabla_{v_2}v_2 = - \frac{k_P}{2}\mathfrak{grad}^2 V_{21} - k_D v_2.
    \end{align}
   
\end{itemize}

Observing that in either case $V_{12}=V_{21}=V$ and that the gradients are negatives of each other, that is, $\mathfrak{grad}^2 V_{21}=-\mathfrak{grad}^1 V_{12} := \mathfrak{grad}V$, we see that the equilibrium conditions for both the dynamical systems is given by 
\begin{align}
    \mathfrak{grad}^1 V_{12} =0.
\end{align}

This implies that at equilibrium,
\begin{align}
    g_{1}^{-1}*g_2\in C.
\end{align}

So, we see that for the two agent case, there are only trivial equilibrium solutions - the consensus case (when $g_1^{-1}*g_2=e$) and the anti-consensus cases $g_{1}^{-1}*g_2=c\in C-\{e\}$. There are no non-trivial equilibria. 
\subsubsection{Line Graph}

For multiple agents, first let  us start with a simple graph where the an agent $i$ is connected to just $i+1$ and $i-1$. (agent 1 is connected only to 2 and agent $n$, only to $n-1$).

The equilibrium condition for such $a_{ij}$ looks as follows:

\begin{align}
    a_{12} \mathfrak{grad}^1_{E_{12}} V_{12} = 0 \nonumber,\\
    a_{21} \mathfrak{grad}^2_{E_{21}} V_{21} + a_{23} \mathfrak{grad}^2_{E_{23}} V_{23}=0 \nonumber, \\
    a_{32} \mathfrak{grad}^3_{E_{32}} V_{32} +  a_{34} \mathfrak{grad}^3_{E_{34}} V_{34}  = 0 \nonumber, \\
    \vdots \nonumber\\
   % a_{n-1,n-2}\mathfrak{grad}^{n-1}_{E_{n-1,n-2}} V_{n-1,n-2} +  a_{n-1,n} \mathfrak{grad}^{n-1}_{E_{n-1,n}}   V_{n-1,n} =0\nonumber\\
    a_{n,n-1} \mathfrak{grad}^{n}_{E_{n,n-1}}   V_{n,n-1} =0.
\end{align}

First, we have $\mathfrak{grad}^1_{E_{12}} V_{12} = 0$ and since $\mathfrak{grad}^2_{E_{21}} V_{21} = -\mathfrak{grad}^1_{E_{12}} V_{12} $, that term in the second equation also vanishes. So, the only remaining term in the second equation is $a_{23} \mathfrak{grad}^2_{E_{23}} V_{23}$ which also has to vanish. Proceeding further, we have 
\begin{align}
 \forall \hspace{1mm} i=2,3,\ldots,n: \hspace{3mm}   &\mathfrak{grad}^{i}_{E_{i,i-1}} V_{i,i-1} =0, \\ 
 \nonumber \Longleftrightarrow \hspace{2mm}  &g_{i-1}^{-1}*g_{i} \in C.
\end{align}

But since $C$ is a subgroup (now we see why that property is crucial), we have
\begin{align}
    g_i^{-1}*g_j&=\overbrace{g_i^{-1}*g_{i+1}}^\text{$\in$ C}*\overbrace{g_{i+1}^{-1}*g_{i+2}}^\text{$\in$ C}*\cdots \nonumber \\ & \cdots *  \overbrace{g_{j-1}^{-1}*g_j}^\text{$\in$ C} .
\end{align}
So, $g_i^{-1}*g_j\in C$ for any $i,j$ since it is a group product of elements in $C$ and $C$, being a subgroup is closed under group operation. So, we have that all pair of agents are C related and the individual gradients $grad^i V(g_j^{-1}*g_i)=0$ vanish. So, the equilibrium consists of only the trivial consensus and anti-consensus points. 

\subsubsection{Trees}
We now consider the case when the graph is connected and is a tree. We  establish once again that in a tree, the equilibria only consists of the trivial consensus and anti-consensus points.

\begin{thm}
When a graph $\mathcal{G}$ is a connected tree, the only solutions that satisfy the equilibrium condition in  \eqref{eqm} are trivial - only the consensus and anti-consensus points.
\end{thm}

\begin{figure}[h]
    \centering
    \includegraphics[scale=0.35]{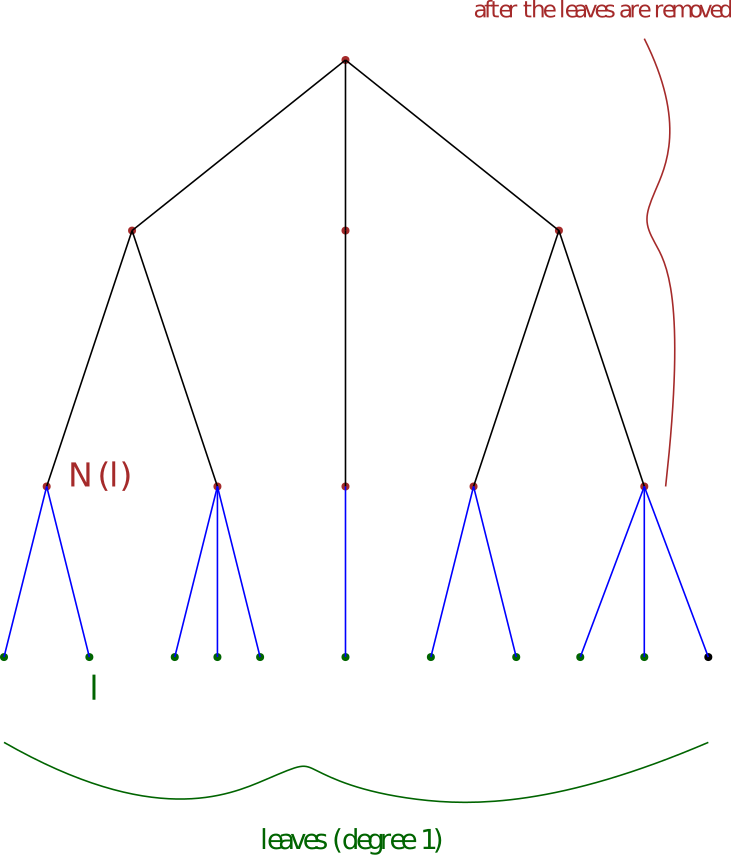}
    \caption{A visualization of a tree with its leaves and its neighbors}
    \label{fig:tree}
\end{figure}

\textbf{Proof:}\\
Consider a connected tree graph, an example of which is shown in Figure \ref{fig:tree}. 
From Theorem \ref{treet}, leaves exist. Let $l$ be a leaf of the tree graph $\mathcal{G}$. Then, it has degree 1 by definition and hence in the dynamics of the agent $l$, there can be only one gradient term. Let the lone neighbor of $l$ be $N(l)$. Then we have the equilibrium condition of agent $l$ as
\begin{align}
    a_{l,N(l)}\hspace{1mm}\mathfrak{grad}^l_{E_{l,N(l)} } V_{l,N(l)} =0.
\end{align}
So we now have that
\begin{align}
    g_{l}^{-1}*g_{N(l)}\in C.
\end{align}
Now since these gradients associated with the leaves $l$ vanish, we can substitute zero in their place and now consider the reduced connected tree that is obtained after removing the leaves and their associated edges. 

Now, one can do the same process for the reduced tree. Consider their leaves. One can then equate their associate gradients to zero as well and move on by removing them.

One can continue this until the reduced tree becomes empty. The reduction will lead to an empty graph in finite steps since in each step, atleast one node is removed (every tree has atleast one leaf and we are removing leaf nodes at each step). So, since the graph ultimately reduces to the empty graph, eventually, every node will be removed as a leaf at some stage. But since the gradients for edges associated with leaf nodes are proven to vanish at each step, we have that the gradient terms for every edge associated with every node vanishes. So, we have 
\begin{align}
    g_k^{-1}*g_l\in C \hspace{2mm} \text{if $(k,l)$ is an edge}.
\end{align}

Since the graph $\mathcal{G}$ is connected, there is a path made of edges, between any two nodes. Let $i,j$ be two arbitrary nodes, with a path $(i,i'),(i,i''),\cdots (i^{''\cdots'},j)$. Then, again since $C$ is a subgroup, we have
\begin{align}
       g_i^{-1}*g_j&=\overbrace{g_i^{-1}*g_{i'}}^\text{$\in$ C}*\overbrace{g_{i'}^{-1}*g_{i''}}^\text{$\in$ C}*\cdots \nonumber \\ & \cdots *  \underbrace{g_{i''\cdots'}^{-1}*g_j}_\text{$\in$ C} .
\end{align}
So, for any pair of agents $(i,j)$, at equilibrium, $g_i^{-1}*g_j\in C$,  since $C$ is a subgroup.

\subsection{Energy and formulation as a Gradient Dynamical System}

We saw that $V$ is a spring-like potential energy function that acts to provide a restoring force that tries to bring together, the pair of agents it connects. With this intuition, let us analyze the behavior of the energy function for the first and second-order Laplacian flow.

Inspired from the Euclidean and circle case, consider the total potential energy function (the $\frac{1}{2}$ in the potential energy term $V_T$ is because each spring is counted twice).
\begin{align}
    V_T (g_1,\ldots,g_n) = \frac{1}{2}\sum_{j,k=1}^n a_{jk}V_{jk} = \sum_{j<k} V_{jk} \label{vt}.
\end{align}

Let us evaluate the gradient of $V_{T}$.

\begin{align}
    grad^i V_T &=\frac{1}{2}\sum_{j,k=1}^n a_{jk} grad^i V_{jk} \\ &= \frac{1}{2}\sum_{i,k=1}^n a_{ik} grad^i V_{ik} + \frac{1}{2}\sum_{j,i=1}^n a_{ji} grad^i V_{ji} \\ &= \sum_{i,k=1}^n a_{ik} grad^iV_{ik} .
\end{align}
We see that the gradient of $V_T$ is the consensus term appearing in the first and second-order dynamics in  \eqref{gfolf}-\eqref{gsolf}. So, they can be rewritten as
\begin{align}
    &\text{First-Order Laplacian Flow:} \nonumber\\
    v_i &= -k_P \mathfrak{grad}^i  \hspace{1mm}  V_T, \label{gfolfg} \\
    &\text{Second-Order Laplacian Flow:} \nonumber\\
    \nabla_{v_i}v_i &= -k_P \mathfrak{grad}^i \hspace{1mm} V_T - k_D  v_i \label{gsolfg}.
\end{align}

So, the equilibria of the systems, as required in \eqref{eqm}, are nothing but the critical points of $V_T$. 
We now state a standard result from gradient dynamical systems as follows:
\begin{thm} \cite{BulloNS}
Consider a first or second-order gradient descent dynamical system in the form
\begin{align}
    \dot{x}&=-k_P grad(F),\\
    \nabla_{\dot{x}}\dot{x} &= -k_P grad(F) - k_D \dot{x}.
\end{align}
Assume that $k_P,k_D>0$ and the sub-level set corresponding to the initial position is compact. Then, the trajectory starting from that initial condition with any initial velocity satisfies the following:
\begin{itemize}
    \item Its velocity goes to zero asymptotically.
    \item Its position that is attained asymptotically is one of the critical points of $F$.
    \item If a critical point has a direction that maximizes $F$, then that point is never locally stable and has an unstable manifold of non-zero dimension.
    
\end{itemize}
\end{thm}

So, we see that if the group $G$ is compact, all trajectories of the first and second-order Laplacian flow converge to critical points of $V_T$ which is the equilibrium of the dynamics as well. So, the equilibrium set of the system is globally asymptotically stable. But it includes, as we saw, consensus points, anti-consensus points and other non-trivial solutions for generic graphs. We can rule out the non-trivial solutions for special graphs as was seen in the previous section - for two agents, or when the graph is a tree. 

\subsection{Local Asymptotic Stability of the Consensus Solution}

We now show that the consensus solution is locally asymptotically stable for both the first and second-order systems and that the anti-consensus solutions are unstable. For the first-order dynamical system in \eqref{gfolf}, the Jacobian of the dynamical system is nothing but the negative Hessian of $V_{T}$ as the vector field is itself the negative gradient of $V_T$. 
Recall that for a smooth vector field $W$  in a Riemannian manifold $G$ with an induced covariant derivative $\nabla$, the \textit{Riemannian Jacobian}  at a point $p$, denoted by $J_p(W):T_pG\rightarrow T_pG$ is a linear map, defined in geometric coordinate invariant terms as 
\begin{align*}
    J_p W (v)=(\nabla_v W)_p.
\end{align*}
When the vector field $W$ is the gradient of a scalar function $F$, putting $W=grad(F)$,  the Jacobian of $grad(F)$ at a point $p$ is called the \textit{Riemannian Hessian of $F$} defined as 
\begin{align*}
    Hess_p F (v) = (\nabla_v grad(F))_p.
\end{align*}
When the manifold is a Lie group, the vector field, the input vector, the gradient vector, all can be left transported to $\mathfrak{g}$ and the connection also can be transported to the Lie algebra and hence the Hessian/Jacobian can be defined as a map from $\mathfrak{g}$ to $\mathfrak{g}$. Denoting the Lie-algebraic Jacobian and Hessian by $\mathfrak{J}$ and $\mathfrak{hess}$, and the Lie algebraic vector field by $\mathfrak{W}$, we have
\begin{align}
  \mathfrak{J}W (v)&=\nabla_v \mathfrak{W},  \\
  \mathfrak{hess}F (v)&= \nabla_v \mathfrak{grad}F.
\end{align}
Let us evaluate the Jacobian matrix of the first-order consensus flow \eqref{gfolfg}, which is also the negative of the Hessian matrix of $V_{TOT}$. Since the configuration space of the system is a product space of $G$, $n$ times, the Jacobian also splits into blocks as $\mathfrak{J}=[J]_{ij}:=-\nabla^j \mathfrak{grad}^i V_{TOT}$ where $\nabla^j$ is the covariant derivative with respect to only the velocity of the $j^{th}$ agent , that is, $v=(0,0,\ldots,v_j,0,\ldots,0)$. The linearity of the Jacobian enables one to split up the Jacobian as block matrices and hence we have 
\begin{align}
    [\mathfrak{J}(v_1,\ldots,v_k,\ldots,v_n)]^i=\sum_k J_{ik}v_k.
\end{align}
We now evaluate the Jacobian, block-by-block.
\begin{lem}
  The block Jacobians are
  \begin{align}
    J_{ii}&=-\sum_{j\neq i} a_{ij}\nabla^i \mathfrak{grad}^i V_{ij} \label{commp1},\\
    J_{ij}&=a_{ij}\nabla^i \mathfrak{grad}^i V_{ij}. \label{commp2}  \end{align}
\end{lem}

\textbf{Proof:}\\

If $i\neq j$, we have
\begin{align}
    J_{ij}=-\nabla^j \sum_k a_{ik}\mathfrak{grad}^i V_{ik} = -a_{ij}\nabla^j\mathfrak{grad}^i V_{ij} . \label{sim}
\end{align}
When $i=j$, the diagonal block evaluates to
\begin{align}
    J_{ii}&=-\nabla^i \sum_{k\neq i} a_{ik}\mathfrak{grad}^i V_{ik}=-\nabla^i \sum_{j\neq i} a_{ij}\mathfrak{grad}^i V_{ij} \nonumber\\
    &= -\sum_{j\neq i} a_{ij}\nabla^i \mathfrak{grad}^i V_{ij}.
\end{align}

We now simplify equation \eqref{sim}.
%Noting that $V_{ij}=V_{ji}$, we have
\begin{align}
    J_{ij}=-a_{ij}\nabla^j\mathfrak{grad}^i V_{ij} = a_{ij}\nabla^j \underbrace{(-\mathfrak{grad}^i V_{ij})}_{grad^j V_{ji} \hspace{1mm}\text{from theorem \ref{oppt}}}.
\end{align}
Recalling the result in Theorem \ref{oppt} that $ \mathfrak{grad}^i V_{ij} = -\mathfrak{grad}^j V_{ji}$,
%and noting that $V_{ij}=V_{ji}$
 we now have

\begin{align}
    J_{ij}=a_{ij}\nabla^j \mathfrak{grad}^j V_{ji}.
\end{align}

We now note that an important feature of the Riemannian Hessian that it is a self-adjoint operator. Hence, it is symmetric as a block matrix. Hence we have $J_{ij}=J_{ji}$. Hence, we have
\begin{align}
    J_{ij}&=a_{ij}\nabla^j \mathfrak{grad}^j V_{ji}=  \nonumber\\
    J_{ji}&=a_{ji}\nabla^i \mathfrak{grad}^i V_{ij}\nonumber\\ &=a_{ij}\nabla^i\mathfrak{grad}^i V_{ij}.
\end{align} 

Once again, putting the entries of $J_{ij}$ and $J_{ii}$ side by side, we see that we have the required result \eqref{commp1}-\eqref{commp2}. 

% \begin{align}
%     J_{ii}&=-\sum_{j\neq i} a_{ij}\nabla^i \mathfrak{grad}^i V_{ij} \label{comp1}\\
%     J_{ij}&=a_{ij}\nabla^i \mathfrak{grad}^i V_{ij}. \label{comp2}  \end{align}
Let us compare  \eqref{commp1}-\eqref{commp2} to the definition of the negative of the Laplacian matrix as follows:
\begin{align}
    -L_{ii}&=-(O-A)_{ii}=-O_{ii}=-\sum_{j\neq i} a_{ij},\\
   - L_{ij}&=-(O-A)_{ij}=(A-O)_{ij}=a_{ij}.
\end{align}
So, we have finally,
\begin{align}
-L_{ii}&=-\sum_{j\neq i}a_{ij} \label{coo1},\\
-L_{ij}&=a_{ij}. \label{coo2}
\end{align}
Now, replacing $a_{ij}$ in  \eqref{coo1}-\eqref{coo2} by $a_{ij}\nabla^i \mathfrak{grad}^i V_{ij}$, we get the Jacobian matrix  \eqref{commp1}-\eqref{commp2}. So, the block Jacobian matrix looks exactly like the negative of the Laplacian matrix except that the weights $a_{ij}$ now are multiplied by a block Hessian $\nabla^i\mathfrak{grad}^i V_{ij}$. 

In the graph theoretic preliminaries, it was mentioned that the Laplacian matrix can be factorised as $L=BDB^T$ where $B$ is the incidence matrix and $D$ is the diagonal matrix consisting of edge weights. With a modified notion of block incidence matrix defined as 
\begin{align*}
    \bar{B}_{ie}:&=+I_{m\times m}, \hspace{2mm} \text{if node $i$ is the source of the edge $e$},\\
    \bar{B}_{ie}:&=-I_{m\times m}, \hspace{2mm} \text{if node $i$ is the sink of the edge $e$},\\
    \bar{B}_{ie}:&=\hspace{1mm}0_{m\times m}, \hspace{2mm} \text{otherwise},\\
\end{align*}
and a modifed block diagonal matrix defined as 
\begin{align*}
    \bar{D}_{ee}=a_{ij}\nabla^i\mathfrak{grad}^i V_{ij},
\end{align*}
we can analogously factorize the Jacobian $\mathfrak{J}$ as 
\begin{align}
 \mathfrak{J}=-\bar{B}\bar{D}\bar{B}^T   .
\end{align}

Since $\mathfrak{J}=-\bar{B}\bar{D}\bar{B}^T$, we are assured the negative semi-definiteness of $\mathfrak{J}$ if we are sure that each block diagonal matrix $a_{ij}\nabla^i \mathfrak{grad}^i V_{ij}$ is positive definite. Since $a_{ij}>0$, we must have that the map $\nabla^i \mathfrak{grad}^i V_{ij}$ be positive definite. We will now show that this is indeed the case as long as $E_{ij}=g_j^{-1}*g_i$ belongs to a small enough neighborhood of  $e$. 
\begin{lem}
  The map $\nabla^i\mathfrak{grad}^i V_{ij}$ is positive definite if $E_{ij}=g_j^{-1}*g_i$ is restricted to a neighboorhood $U$ around $e$.
\end{lem}

\textbf{Proof:}\\ 
\begin{align*}
    \nabla^i \mathfrak{grad}^i V_{ij} &= \nabla^i \mathfrak{grad}^i V(g_j^{-1}*g_i)\\&=\nabla^i \mathfrak{grad}_{g_j^{-1}*g_i} V\\&=\mathfrak{hess}(V)_{g_j^{-1}*g_i}. \label{lu}
\end{align*}

But since $V$ is a polar Morse function, the Hessian at identity $\mathfrak{hess}_e$ is positive definite as $e$ is the global minimum. Now, since $\mathfrak{hess}$ is continuous (as $V$ is smooth), and the set of positive definite matrices is open, we have that $\mathfrak{hess}$ is positive definite in a neighborhood $U$ of $e$ as well. Hence, the result ensues. 
\begin{thm}
  The consensus equilibrium is locally exponentially stable for any connected symmetric graph for both the first and second order dynamics.
\end{thm}
\textbf{Proof:}\\

So, the matrix $\bar{D}$ is positive definite if $E_{ij}\in U \hspace{2mm} \forall i,j$. But, we are concerned about the consensus equilibrium where $E_{ij}=e\hspace{2mm}\forall i,j$ and hence the matrix $\bar{D}$ is trivially positive definite. Hence, $\mathfrak{J}$ is negative semi-definite. Since the  the graph $\mathcal{G}$ is connected, the null space of $\mathfrak{J}$ will consist only of consensus velocities of the form $(v,v,\ldots,v)$ where $v\in \mathfrak{g}$, that is $ker(\mathfrak{J})=\{(v,v,\ldots,v)\hspace{2mm} v\in \mathfrak{g}\}$. But these consensus velocities span the tangent space to the consensus manifold $M_C:=\{(g,g,\ldots,g): g\in G\}$. Also, the consensus manifold is invariant (since agents stay in consensus once they start at consensus). So, $M_C$ is an invariant submanifold spanning $ker(\mathfrak{J})$. So, the uniqueness of the center manifold  (by center manifold emergence theorem), it is the centre manifold and hence in a neighborhood, any solution exponentially converges to a solution in $M_C$ and hence consensus is locally exponentially stable.

The local asymptotic stability of the consensus solution in second-order flow follows from the standard result that the asymptotic behavior of first and second-order gradient systems are equivalent with $k_P,k_D>0$. It follows easily from the fact that the spectrum of a matrix $A$ lies in the left/right half plane if and only if the spectrum of the matrix $\begin{bmatrix} 0 & I \\ -A & -k_D I \end{bmatrix}$  lies in the left/right half plane for any $k_D>0$. Applying this result to the Jacobian of the second-order system yields local exponential stability of the consensus manifold there as well.

\begin{thm}
The other trivial anti-consensus equilibria where all the inter-agent errors are not necessarily $e$ but in $C$ are unstable.
  
\end{thm}
\textbf{Proof:}\\
    To prove this, let us choose an agent pair $(i,j)$ where $E_{ij}=c  \neq e$ where $c\in C$. 
    %Let us choose a direction $(v_1,v_2,\ldots,v_n)$ such that $v_i=v_j=v\in \mathfrak{g}$ and rest of the velocities are zero - that is $v=(0,0,\ldots,\underbrace{v}_{i},\ldots,\underbrace{v}_{j},\ldots,0)$.  
    Then, in $V=\frac{1}{2}\sum_{i,j=1}^n a_{ij}V_{ij}$, if we vary only $g_i,g_j$,  $V_{ij}$ is a saddle with at least one negative eigen value at $E_{ij}=c\in C$ and hence there is a direction in which it is maximum. So, there is an unstable direction and hence the trivial equilibrium is not attracting if it is not a consensus equilibrium.

As a Corollary, since a gradient system on a compact manifold can only converge to an equilibrium asymptotically, we have that the consensus equilibrium has an almost global domain-of-attraction if all other equilibria are unstable (since it is the only attracting equilibrium). 
\section{Synchronization on Compact Lie Groups}

We can also generalize the synchronization protocol from oscillators to Lie groups. Compare the consensus and synchronization protocols for oscillators as follows:
\begin{align}
    &\text{Consensus:} \nonumber\\ 
    \dot{\theta}_i &= -k_P\sum_{j=1}^na_{ij}\sin(\theta_i-\theta_j), \\
    &\text{Synchronization:} \nonumber \\ 
     \dot{\theta}_i &= \omega_i-k_P\sum_{j=1}^na_{ij}\sin(\theta_i-\theta_j).
\end{align}
The only extra term in the synchronization protocol  (compared to the consensus dynamics), is an extra angular velocity $\omega_i$. So, with this intuition, we propose a synchronization protocol in Lie groups as

\begin{align}
      v_i = \mathfrak{w}_i - k_P  \sum_{j=1}^n a_{ij} \hspace{1mm} \mathfrak{grad}^i_{E_{ij}} V_{ij}. \label{sync}
\end{align}

Here, $\mathfrak{w}_i$ is now a constant velocity in the Lie algebra, and is a generalization of the intrinsic natural frequency of the oscillator $\omega_i$ in $\mathbb{S}^1$. In the oscillator case, without the coupling, each oscillator would exhibit a periodic motion with angular velocity $\omega_i$. In this case, when the coupling gain $k_P$ is zero, each $i^{th}$ agent travels with a constant left  velocity $\mathfrak{w}_i$ (or the trajectory of the agent is an integral curve of the left invariant vector field generated by $\mathfrak{w}_i$). In the case of an oscillator that evolves on a circle, such a curve with constant velocity $\omega_i$ is periodic. In case of tori that are Cartesian products of circles, the curves will be quasi-periodic - periodic in each angular component. We now establish the same result for general compact Lie groups. That is, when a Lie group $G$ is compact, their integral curves of left invariant vector fields are also quasi-periodic. 

\begin{thm}
The integral curves of left invariant vector fields on a compact Lie group are quasi-periodic.
\end{thm}

\textbf{Proof:}\\: As is standard from representation theory, every compact Lie group has a faithful matrix representation by unitary matrices of finite-dimension. In other words, every connected and compact Lie group is isomorphic to a matrix Lie subgroup of $U(n)$. So, since the Lie algebra of $U(n)$ is $\mathfrak{u}(n)$ (space of $n\times n$ skew Hermitian matrices), the left invariant vector field on a matrix Lie group of orthogonal matrices can be written as
\begin{align*}
    \dot{X}&=X\mathfrak{w}, \hspace{2mm}\text{where} \hspace{1mm} \mathfrak{w}\in \mathfrak{u}(n)\\
        \text{Integral curve:} \hspace{2mm}
    X(t)&= e^{\mathfrak{w}t}X(0).
\end{align*}
Now since $\mathfrak{w}$ is a skew Hermitian matix, it has purely imaginary eigen values $\pm i \omega_k$. When $\mathfrak{w}$ is diagonalized, and the ODE is rendered in diagonalized coordinates, we hence have $\frac{d}{dt} x_i = \pm i\omega_i x_i$ which integrates to give $x_i(t)=e^{\pm i\omega_i t}x_i(0)$ which is periodic in each $x_i$. So, we see that the integral curve is resolvable into periodic motions and hence is quasi-periodic. 

When all the $\mathfrak{w}_i=0$, then the equation reduces to the standard Laplacian flow with just the consensus term, as usual.

\subsection{Synchronization notions}
So, now we have that the agents have a natural tendency to move at some constant left velocity in addition to trying to achieve consensus.
Recalling the synchronization protocol again
\begin{align*}
     v_i = \mathfrak{w}_i - k_P  \sum_{j=1}^n a_{ij} \hspace{1mm} \mathfrak{grad}^i_{E_{ij}} V_{ij} ,
\end{align*}
we are curious as to what ensues, as a result of these two opposing tendencies for each of the agent. 

We now extend the definitions from the oscillator case to the generic case. Let $U$ be a given open neighborhood of identity that is symmetric under inversion ($U=U^{-1}$) (such neighborhoods can easily be constructed as if $U$ is any other neighborhood of $e$, $U\cap U^{-1}$ satisfies this requirement). 
\begin{defn}
Trajectories  $g_i(t)$ evolving on a compact Lie group $G$ are said to be
\begin{itemize}
 \item \textbf{position synchronized} if  $g_i(t)=g_j(t)$, $\forall \hspace{1mm}t$, $\forall \hspace{1mm} i,j$.
 \item  \textbf{left velocity synchronized} if their left velocities coincide , that is $g_i^{-1}(t)*\dot{g}_i(t)=g_j^{-1}*\dot{g}_j(t)$ $\forall \hspace{1mm} t$ $\forall \hspace{1mm}  i,j$.
 \item \textbf{$U$-cohesive} if $g_j^{-1}(t)*g_i(t) \in U$ $\forall \hspace{1mm} t$ ,$\forall \hspace{1mm} i,j$.
\end{itemize}
\end{defn}

With this , we prove a couple of simple results.

\begin{thm} Consider the dynamical system in \eqref{sync}.
\begin{enumerate}
  
    \item Phase synchronization is possible if and only if all the  $\mathfrak{w}_i$ coincide (agents share a common natural left velocity)
    \item If velocity synchronization occurs in the dynamics \eqref{sync}, then the synchronous velocity attained (denoted by $\mathfrak{w}_S$) can only be equal to the average of the individual natural left velocities.
    \begin{align}
        \mathfrak{w}_S = \frac{1}{n} \sum_{i=1}^n \mathfrak{w}_i  . \label{trt}
        \end{align}
    \end{enumerate}
\end{thm}

\textbf{Proof:}\\
\begin{enumerate}
    \item If $g_i(t)=g_j(t)$ (which implies $E_{ij}=e$), then differentiating, $\dot{g}_i(t)=\dot{g}_j(t)$ which gives $v_i=v_j$. Substituting the dynamics for $v_i$ in \eqref{sync} and noting that all gradients vanish (all errors being $e$ and gradient of $V$ vanishes at $e$), we have that $v_i=\mathfrak{w}_i+0=\mathfrak{w}_i$. Since $v_i=v_j$, $\mathfrak{w}_i=\mathfrak{w}_j$. 
    \item Summing over $i$ in \eqref{sync} , we have \begin{align}
    \mathfrak{w}_S &= \frac{1}{n}\sum_{i=1}^n \mathfrak{w}_S\nonumber \\
        &=\frac{1}{n}\sum_{i=1}^n v_i=\frac{1}{n}\sum_{i=1}^n \mathfrak{w}_i - \frac{1}{n}k_P \overbrace{\sum_{i=1}^n\sum_{j=1}^n \mathfrak{grad}^i_{E_{ij}} V_{ij}}^\text{add to 0} \label{cn}.
    \end{align}
\end{enumerate}
But by Theorem \eqref{opp} in \ref{oppt}, we have $ \mathfrak{grad}^i V_{ij} = -\mathfrak{grad}^j V_{ji}$ and hence the gradient terms in \eqref{cn} add to zero pairwise. Hence, \eqref{trt} ensues. 

\textbf{Note:} When the agents are all moving with a constant left-synchronous velocity $\mathfrak{w}_S$, then their flow is given by $g_i(t)=\mathrm{e}^{\mathfrak{w}_St} g_i(0)$ and hence their left-invariant errors $E_{ij}=g_j^{-1}(t)g_i(t)=g_j(0)^{-1}g_i(0)$ remain constant. So, the arguments of $V_{ij}$ remain constant. So, $E_{ij}$ are constants for all $i,j$. 

\subsection{A necessary condition for the existence of synchronization solution}

Next, we need to look deeply, what are some necessary and sufficient conditions for which the synchronous solution exists. The following are some investigations. We first prove some elementary results. 

\begin{thm}
Topological result :
 Consider the gradient function $\mathfrak{grad}: G \rightarrow \mathfrak{g}$. Then, the image of $G$ under the function $\mathfrak{grad}$ is a compact (hence bounded) subset of $\mathfrak{g}$, containing a neighborhood $U'$ of origin. It is also possible to choose this neighborhood $U'$ such that the function $\mathfrak{grad}_g V$ is locally injective and surjective (hence a local diffeomorphism) in that neighborhood.  \label{lam}
 \end{thm}
 
 \textbf{Proof:}\\ 
 
 Since $G$ is a compact and connected Lie group by assumption and $\mathfrak{grad}V$ is smooth (and hence continuous), the image set $\mathfrak{grad}V (G)$ will also be compact and connected as continuous functions preserve compactness and connectedness. As a Corollary, $\mathfrak{grad}V$ is bounded in $M$ and hence its norm has a supremum that is actually attained in $M$. 
 
\begin{figure}[h]
     \centering
     \includegraphics[scale=0.35]{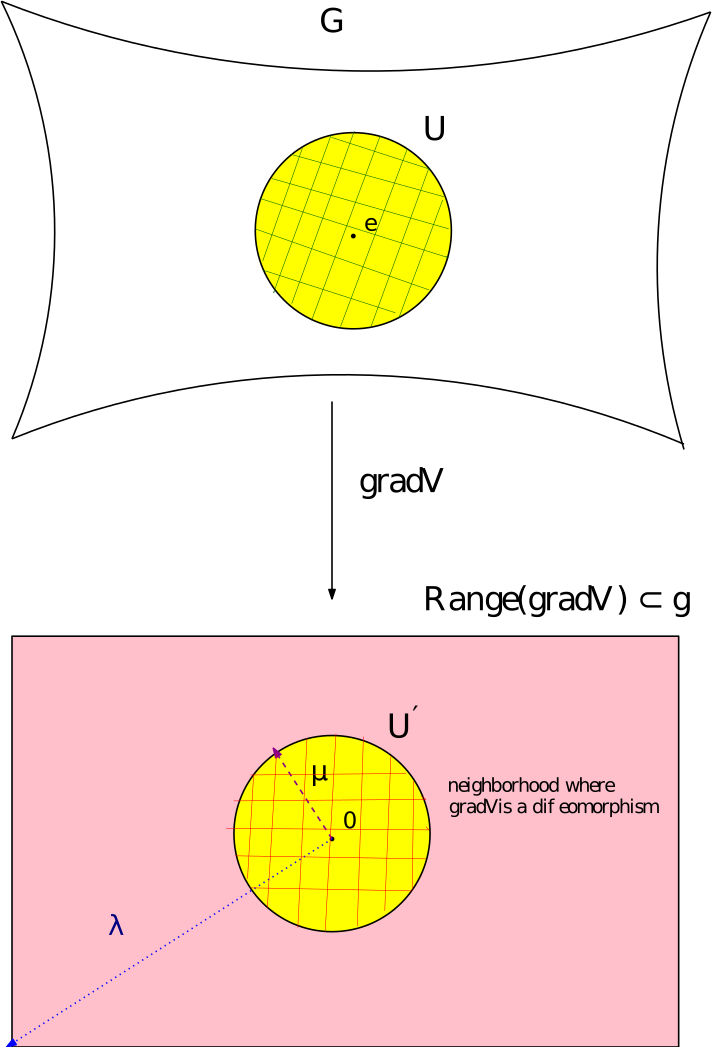}
     \caption{Visualization of the topology of $\mathfrak{grad}:G\rightarrow \mathfrak{g}$}
     \label{fig:fig2}
 \end{figure}

 Since $e$ is the global minimum of $V$, it is also a local minimum and choose a local coordinate system around $e$ so that $F(x_1,\ldots,x_m)=x_1^2+x_2^2+\cdots+x_m^2$. (By Theorem \ref{mnf}). Then the differential of $F$ in this coordinate system is $\frac{\partial F}{\partial x_i}=2x_i$ and hence $grad(F(x))=2x$. Hence, locally the gradient map is scaled identity in this local coordinate system for $G$, and hence is injective, surjective and smooth. This makes $\mathfrak{grad}V$, a local diffeomorphism around $e$ (by inverse function theorem). But local injectivity and surjectivity are coordinate independent facts  and hence generalize to any coordinate system for $G$ and $\mathfrak{g}$. Since $\mathfrak{grad}V(e)=0$, the identity $e$ gets mapped to origin in $\mathfrak{g}$  and hence a coordinate neighborhood around $e$ gets mapped to a neighborhood of origin in $\mathfrak{g}$ since the map is a local diffeomorphism. Restricting $\mathfrak{grad}V$ to this coordinate neighborhood of $e$ on $G$ and considering its image in $\mathfrak{g}$, we get the required result. The result is visualized in Figure \ref{fig:fig2}. 
 
 \textbf{NOTE:} For the case of the circle (that is of interest to oscillator synchronization) where $V=1-\cos\theta$, and $grad(V)=-\sin\theta$, the neighborhood around identity is nothing but $(-\frac{\pi}{2},\frac{\pi}{2})$ as the sine function $\sin:(-\frac{\pi}{2},\frac{\pi}{2})\rightarrow \mathbb{R}$ is injective and hence $\sin:(-\frac{\pi}{2},\frac{\pi}{2})\rightarrow (-1,1)$ is a local diffeomorphism. And $\mathrm{sup}_{\theta} ||\mathfrak{grad}(V)|| = \mathrm{sup}_{\theta}|\sin\theta|=1$ for the circle.

 \begin{thm}
   If $\lambda:=\mathrm{sup}_G\hspace{1mm} ||\mathfrak{grad} \hspace{1mm} V||$ (it exists by Theorem \ref{lam}) , and $||\mathfrak{w}_i-\mathfrak{w}_S||>deg(i)k_P\lambda $ for some $i$, then there can be no velocity synchronous solution for the system. (the agents cannot attain velocity synchrony if the deviation from synchronous velocity is large enough compared to the coupling strength)

\end{thm}

\textbf{Proof:}\\
Substracting $\mathfrak{w}_S$ from the synchronization dynamics \eqref{sync}, we get
\begin{align}
      v_i - \mathfrak{w}_S = (\mathfrak{w}_i - \mathfrak{w}_S) - k_P  \sum_{j=1}^n a_{ij} \hspace{1mm} \mathfrak{grad}^i_{E_{ij}} V_{ij} .\label{dsync}
\end{align}

The LHS of \eqref{dsync} has to vanish if the agents are in synchrony. Hence we must have
\begin{align}
       (\mathfrak{w}_i - \mathfrak{w}_S) =  k_P  \sum_{j=1}^n a_{ij} \hspace{1mm} \mathfrak{grad}^i_{E_{ij}} V_{ij} \label{ddsync}.
\end{align}

But since $0\leq a_{ij}\leq 1$, we have by triangle inequality, 
\begin{align}
    ||\mathfrak{w}_i - \mathfrak{w}_S||= \bigg|\bigg| k_P  \sum_{j=1}^n a_{ij} \hspace{1mm} \mathfrak{grad}^i_{E_{ij}} V_{ij} \bigg|\bigg| \nonumber\\ \leq k_P \sum_{j=1}^n  a_{ij} \hspace{1mm} ||\mathfrak{grad}^i_{E_{ij}} V_{ij}||  
     &\leq k_P \lambda \underbrace{\sum_{j=1}^na_{ij}}_{\text{=deg(i)}} 
.\end{align}

So, when the agents are synchronous, we have that $||\mathfrak{w}_i - \mathfrak{w}_S||\leq deg(i)k_P \lambda$. So, synchrony cannot happen if $||\mathfrak{w}_i-\mathfrak{w}_S||>deg(i)k_P\lambda$.  

In particular, if $||\mathfrak{w}_i-\mathfrak{w}_S|| > k_P\lambda \mathrm{max}_{i}deg(i)$, then velocity synchronization cannot occur. So, if the maximum deviation from synchronous speed of an agent exceeds a large enough value, then the agents cannot achieve velocity synchronization.

\subsection{A sufficient condition for the existence of synchronization solution}
Now, a sufficient condition for the existence of velocity synchronous solution is investigated. 

\subsubsection{Two Agent Case}
Let us consider first, the case of a two agents to get a geometric intuition. In case of two agents, with adjacency matrix elements chosen as $a_{12}=a_{21}=\frac{1}{2}$, we have the dynamics as
\begin{align}
    v_1-\mathfrak{w}_S = \mathfrak{w}_1-\mathfrak{w}_S - k_P \frac{1}{2} \mathfrak{grad}^1 V_{12},\\
    v_2-\mathfrak{w}_S = \mathfrak{w}_2-\mathfrak{w}_S - k_P \frac{1}{2} \mathfrak{grad}^2 V_{21}.
\end{align}
Taking $\mathfrak{grad}^1 V_{12}=-\mathfrak{grad}^2 V_{21}=\xi$, we get
\begin{align}
    v_1-\mathfrak{w}_S = \mathfrak{w}_1-\frac{1}{2}k_P \xi \label{nn1},\\
    v_2-\mathfrak{w}_S = \mathfrak{w}_2 +\frac{1}{2} k_P \xi \label{nn2}.
\end{align}
Adding and subtracting the two equations, we get
\begin{align}
    \frac{1}{2}(v_1+v_2)=\frac{1}{2}(\mathfrak{w}_1+\mathfrak{w}_2),\\
    v_1-v_2 = (\mathfrak{w}_1-\mathfrak{w}_2)-k_p\xi.
\end{align}

So, we have that the velocity synchronization happens only if the following equation has a solution.
\begin{align}
    (\mathfrak{w}_1-\mathfrak{w}_2)=k_P\xi \label{yh}.
\end{align}

Let us now define another scalar quantity $\mu\leq\lambda$ as follows. Let $\mu$ be the radius of the largest ball that is contained in $S$, which is a neighborhood of the origin in $\mathfrak{g}$ where $\mathfrak{grad} V$ is injective and surjective. Then, this means that given a vector $\xi$ with $||\xi||<\mu$, there exists a  unique $g\in U$ such that $\mathfrak{grad}_g V = \xi$.  Let us search for synchronous solutions where the inter-agent error is in this gradient-bijective neighborhood $U$, that is $g_{1}^{-1}*g_2\in U$ .  Due to this restriction, we have that that, given a vector $\xi\in \mathfrak{g}$ with $||\xi||<\mu$,  there exists  a unique $E_{12}=g_1^{-1}*g_2\in U$ such that $\mathfrak{grad}_{E_{12}}V=\xi$.

Returning to the two-agent dynamical system, if $||\mathfrak{w}_1-\mathfrak{w}_2||<k_P\mu$, then \eqref{yh} implies that $||\xi||<\mu$ which is sufficient for the existence of a solution as $||\xi||$ can be attained as a gradient at some point in the group. So, at least for the two agent case, we have that if the  spread in the natural velocities  is sufficiently smaller in magnitude than the coupling strength, then there exists a synchronous solution.

\subsubsection{Star Graph}
A star is a tree where the first node is connected to all $n-1$ nodes and has degree $n-1$ with all the other nodes connected to only the first node. The first node acts as a leader which is connected to all other agents and the other agents following only the leader. In other words, the adjacency matrix is such that $a_{12}=a_{13}=\cdots=a_{1n}=a_{21}=a_{31}=\cdots=a_{n1}=1$ and all other weights zero.
% It is a standard result in spectral theory that the second smallest eigen value $\lambda_2(L)$ of the Laplacian of a star is unity. We have that $deg(1)=n-1$ and $deg(i)=1$ for other $i$.

The equilibrium conditions for the system are:
\begin{align}
 \mathfrak{w}_1-\mathfrak{w}_S&=k_P\sum_{j\neq1} \mathfrak{grad}^1 V_{1j} \label{rnn},\\
     \mathfrak{w}_i-\mathfrak{w}_S&= k_P\mathfrak{grad}^i V_{i1} .
\end{align}

If $||\mathfrak{w}_i-\mathfrak{w}_S||\leq k_P\mu$ for $i\geq2$, then $||\mathfrak{grad}^iV_{i1}||<\mu$ and hence we have a solution for $\mathfrak{w}_i-\mathfrak{w}_S=\frac{1}{k_P}grad_iV_{i1}$. 
Note that once again, we are looking for a synchronous solution where the inter-agent errors $E_{ij}$ are in the bijective neighborhood. i.e. $||\mathfrak{grad}_{E_{ij}}V|| = ||\mathfrak{grad}^i V_{ij}||<\mu.$
Putting $grad^1V_{1j}=-grad^j V_{j1}=-\frac{1}{k_P}(\mathfrak{w}_j-\mathfrak{w}_S)$ in \eqref{rnn}, we get
\begin{align}
    \mathfrak{w}_1-\mathfrak{w}_S&=\sum_{j\neq 1} (\mathfrak{w}_S - \mathfrak{w}_j)\\
    \Longrightarrow n\mathfrak{w}_S&=\mathfrak{w}_1+\sum_{j\neq1}\mathfrak{w}_j=n\mathfrak{w}_S,
\end{align}
which is trivially satisfied as $\mathfrak{w}_S:=\frac{\sum_{j=1}^n\mathfrak{w}_j}{n}$.

So, for the star graph as well, we see that if $\mathfrak{w}_i-\mathfrak{w}_S$ is sufficiently small enough, existence of synchronous solution is guaranteed.

\subsubsection{Line Graph}

Now, let us consider a line graph with weights chosen that $a_{i,i+1}=a_{i+1,i}=1$. 

The synchronization condition in \eqref{ddsync} then gives
\begin{align*}
    \mathfrak{w}_1-\mathfrak{w}_S&= k_P \mathfrak{grad}^1 V_{12},\\
    \mathfrak{w}_2-\mathfrak{w}_S&= k_P (\mathfrak{grad}^2 V_{21} + \mathfrak{grad}^2 V_{23} ), \\
    \mathfrak{w}_3-\mathfrak{w}_S&= k_P (\mathfrak{grad}^3 V_{32} + \mathfrak{grad}^3 V_{34} ), \\
    \vdots \\
     \mathfrak{w}_n-\mathfrak{w}_S&= k_P \mathfrak{grad}^n V_{n,n-1}.
\end{align*}
% \begin{thm}
%   If for each $i$, $||\mathfrak{w}_i-\mathfrak{w}_s||<k_P\mu$, then there exists a synchronous solution for the line graph.
% \end{thm}

% \textbf{Proof:}\\
If $||\mathfrak{w}_1-\mathfrak{w}_S||<\frac{1}{2}k_P\mu$, then $||\mathfrak{grad}_1 V_{12}||<\frac{\mu}{2}\leq \mu$, and hence there is a relative configuration $E_{12}$ such that $\mathfrak{grad}_1V_{12}$ equals $\frac{\mathfrak{w}_1-\mathfrak{w}_S}{k_P}$. 

Having decided the relative configuration of agent 2 with respect to agent 1, let us move to the next equation. 
Next, substituting this, we get to the second equation in the system which is
\begin{align*}
    \mathfrak{w}_2-\mathfrak{w}_S&= k_P (\mathfrak{grad}^2 V_{21} + \mathfrak{grad}^2 V_{23} ) \\
    &= k_P (-\overbrace{\mathfrak{grad}^1 V_{12}}^\text{already determined} + \mathfrak{grad}^2 V_{23} )\\
   \end{align*}.
Hence we have
\begin{align}
    k_P\mathfrak{grad}^2 V_{23}= (\mathfrak{w}_2-\mathfrak{w}_S) + k_P \mathfrak{grad}^1 V_{12}.
\end{align}

For the existence of a solution, we must have $||\mathfrak{w}_2-\mathfrak{w}_S||\leq \frac{k_P\mu}{2}$ as well. If it holds, then $||\mathfrak{grad}^2 V_{23}||\leq \mu$ (by triangle inequality) and hence there exists a relative configuration $E_{23}$. Proceeding further, we see that if $||\mathfrak{w}_i-\mathfrak{w}_S||\leq \frac{k_P\mu}{2}$ for $i=3,\cdots,n-1$, we determine $k_Pgrad^i V_{i,i+1}=(\mathfrak{w}_i-\mathfrak{w}_S) + k_P \mathfrak{grad}_{i-1}V_{i-1,i}$. Hence, by induction, we have
\begin{align}
    k_P  \mathfrak{grad}^iV_{i,i+1}=\sum_{j=1}^i (\mathfrak{w}_j-\mathfrak{w}_S).
\end{align}

With this, we come to the last equation 
\begin{align}
    \mathfrak{w}_n-\mathfrak{w}_S &= k_P grad^n V_{n,n-1} \\&= -k_P grad^{n-1} V_{n-1,n} = - \sum_{j=1}^{n-1} (\mathfrak{w}_j-\mathfrak{w}_S) ,
\end{align}
which is trivially satisfied as $\mathfrak{w}_S:=\frac{1}{n}\sum_{j=1}^n\mathfrak{w}_j$

So, we  get the existence of velocity synchronous solution when the magnitude of deviations are small enough, for the line graph as well.

\subsubsection{Trees}
With the same logic as in consensus, one can generalize these to trees as well. Assuming $||\mathfrak{w}_i-\mathfrak{w}_S||\leq \frac{1}{\mathrm{max}_i deg(i)}k_P\mu$ and  solving first for the leaf nodes and going further down, we can verify that synchronous solution is achieved. This is proved by induction on the number of nodes of the tree. When $n=2$, we have already shown. Let us assume that for a tree of $n-1$ nodes or lower, if  $||\mathfrak{w}_i-\mathfrak{w}_S||\leq \frac{1}{max_ideg(i)}k_P\mu$, then a synchronous solution exists. With this, consider a tree with $n$ nodes. It has aleast one leaf node with degree 1. Without loss of generality, label it the last node $n$ and its lone neighbor as $n-1$. Then, we have at equilibrium,
\begin{align}
    \mathfrak{w}_n-\mathfrak{w}_S = k_P \mathfrak{grad}^{n} V_{n,n-1}.
\end{align}

Since $||\mathfrak{w}_n-\mathfrak{w}_S||<\frac{1}{max_i deg(i)}k_P\mu\leq k_P\mu$, a solution exists for $\mathfrak{grad}_n V_{n,n-1}$. Now, remove the corresponding leaf and the edge to get another connected tree of $n-1$ edges. Wherever one sees $\mathfrak{grad}^{n-1} V_{n-1,n}$, we know that it is equal to $-\mathfrak{grad}^n V_{n,n-1}$ and we know that its magnitude is less than or equal to $\frac{1}{max_i deg(i)}$. Combine it with the term $\mathfrak{w}_n-\mathfrak{w}_S$ and proceeding further using induction, we get the result.

\subsection{Counting argument for the existence of a unique cohesive synchronous solution for arbitrary graphs}

In this section, we give a simple counting argument to support the existence of a synchronous solution when $||\mathfrak{w}_i-\mathfrak{w}_S||$ is small enough in magnitude. 

The system of equations that we have to get a synchronous solution is
\begin{align}
    \mathfrak{w}_i-\mathfrak{w}_S = \sum_{j=1}^n a_{ij}\mathfrak{grad}^i V_{ij} \label{label}.
\end{align}

The important thing is that not all the gradients in the right hand side of \eqref{label} are independent. As an example, let us say that the gradients $\mathfrak{grad}^2 V_{12} =\mathfrak{grad}_{E_{12}}V= \zeta_{12}$ and $\mathfrak{grad}^3V_{23}=\mathfrak{grad}_{E_{23}}=\zeta_{23}$ are known, due to the local bijectivity of the gradient ensures that the errors $E_{12},E_{23}$ are uniquely fixed in this neighborhood of $e$. But once $E_{12}.E_{23}$ are known, it fixes $E_{13}$ as 
$E_{13}=g_1^{-1}*g_3=(g_1^{-1}*g_2)*(g_2^{-1}*g_3)=E_{12}*E_{23}$. So, in a similar manner, once the inter-agent positions $E_{i,i+1}$ are chosen such that gradients $\mathfrak{grad}^i \mathfrak{grad} V_{i,i+1}$ are fixed, all other inter-agent positions and their relative gradients are fixed. Since $i$ runs from $1$ to $n-1$, only $n-1$ gradient terms are independent. So, number of independent unknowns is $n-1$.

How many independent equations do we have? Equations in \eqref{label} might look like $n$ equations but we will soon verify that only $n-1$ of them are independent. This follows from summing the $n$ equations in \eqref{label}.

Doing this, we get
\begin{align}
    \mathfrak{w}_i-\mathfrak{w}_S &= \sum_{j=1}^n a_{ij}\mathfrak{grad}^i V_{ij}\\
    \underbrace{\Longrightarrow \sum_{i=1}^n ( \mathfrak{w}_i-\mathfrak{w}_S)}_{\text{ add to 0 as $n\mathfrak{w}_S=\sum_{i=1}^n \mathfrak{w}_i$ } } &= \underbrace{\sum_{i=1}^n \sum_{j=1}^n a_{ij}\mathfrak{grad}^i V_{ij}}_{\text{add to 0}}.
 \end{align}

So, the $n$ equations are linearly dependent as they all add to zero. This tells that any one equation can be written in terms of the other $n-1$ equations. So, there are only $n-1$ independent equations which matches the number of independent unknowns (the $n-1$ gradients that was shown before). So, this counting argument dictates that there exist a unique solution for the gradients if their magnitudes are low enough to lie in the locally bijective neighborhood.   In particular, since a synchronous solution exists when $\mathfrak{w}_i=\mathfrak{w}_S$ for all $i$ (identical agents - consensus is equivalent to synchronization), for small enough magnitude of $||\mathfrak{w}_i-\mathfrak{w}_S||$, there must still exist a solution to this system of equations. 
% But the gradients $\mathfrak{grad}^n V_{n,j}$ are negatives of the gradients $\mathfrak{grad}^J V_{j,n}$ all of which are present in the remaining $n-1$ equations when $i$ runs from $1$ to $n-1$. Since $a_{nj}=a_{jn}$ (symmetry of the adjacency matrix), the gradients can be written from the previous $n-1$ equations as 
\subsection{Local Asymptotic Stability of Synchronization} 

Finally, we investigate the important question. If a synchronous solution exists in the gradient-injective neighborhood, is it indeed locally exponentially stable? Let us begin by proving some elementary results that would help in establishing a positive answer to the question.

The  dynamics of the entire system of agents can be cast in the form as 
\begin{align}
    v = \mathfrak{w} - k_P\mathfrak{grad}V_{T},
\end{align}
where $v=(v_1,\cdots,v_n)$ and $\mathfrak{w}=(\mathfrak{w}_1,\ldots,\mathfrak{w}_n)$. So, we see that the vector field is addition of two vector fields - a constant left-invariant vector field $\mathfrak{w}$ and a gradient field. In the consensus case, we had only the gradient term. Now, we have an additional drift velocity that is constant (when viewed in the Lie algebra by left translation).

Hence the Jacobian of the vector field is the sum of two Jacobians - the Jacobian of the left invariant vector field $\mathfrak{w}$ and the Jacoobian of the negative gradient of $V_T$ which is the negative Hessian of $V_T$.

We saw from the corollary in \eqref{skss} that the map $L_Y(X):=\nabla_X Y$ is skew symmetric if $Y$ is a left invariant vector field! Replacing $Y$ by $\mathfrak{w}$ (which is also a left-invariant vector field as they are constant left-velocities), we get that the Jacobian of $W$  that is also defined as $\mathfrak{J}\mathfrak{w}(v)=\nabla_v \mathfrak{w}$ is also skew-symmetric. We already saw in the linear algebraic result in the preliminaries that the addition of a skew-symmetric operator to an already stable symmetric operator does not affect the stability property. So, we can investigate the stability of the synchronization system by just investigating the stability of the symmetric part of the Jacobian which is the negative Hessian of $V_T$, thereby ignoring the influence of the drift $\mathfrak{w}$ completely. But the investigation of stability of the negative Hessian of $V_T$ already has been carried out in the consensus case with $\mathfrak{w}=0$. So, those results are transferable now. 

We look for synchronization solution where all the inter-agent errors are in  the gradient-bijective neighborhood $U$, which is  $g_{i}^{-1}*g_j\in U \hspace{2mm} \forall i,j$. A small result that gives insight into the analysis follows:

\begin{thm}
  If $g\in U$, then $\mathfrak{J}\mathfrak{grad}_gV=\mathfrak{hess}_g V$ is always positive definite for all $g\in U$. 
\end{thm}
\textbf{Proof:}\\
Note that when $g\in U$, $\mathfrak{grad}_gV$ is bijective. Therefore, its Jacobian also has to be invertible and always has to be full rank for all $p$. But the Jacobian of a gradient is the Hessian. So, the Hessian has to be full rank for all points in $U$. So, the eigen values of the Hessian have to be non-zero and real (the eigen values of a symmetric matrix like Hessian is always  real). The Hessian at identity $e$ is positive definite as $e$ is the global minimum of the polar Morse function $V$. So, all the eigen values of the Hessian at $e$ are positive. Since the eigen values are continuous functions of the matrices and the Hessian is continuous and smooth as well, they continuously vary in $U$. But they all have to be positive. Even if one of the eigen values at some $p\in U$ were negative, we can have a path joining $e$ to $p$ ($U$ is path connected). Since the eigen values are continous function in the curve and change from being positive at $e$ to negative at $p$, it would have to attain the value zero at some point in the middle, say $p'\in U$. So, the Jacobian of the gradient of $V$ at $p'$ (which is the Hessian of $V$ at $p'$) would have a zero eigen value. This would make the Jacobian non-invertible  at $p'\in U$ and hence contradict the fact that it is full rank and invertible at all points in $U$.

\begin{thm}
\label{previs}
  The synchronous solution where all the agents are phase-cohesive with respect to the gradient-bijective neighborhood $U$ around $e$ - i.e. $\forall i,j \hspace{2mm} g_i^{-1}*g_j\in U$, is always locally exponentially stable.
\end{thm}
\textbf{Proof:}\\
The Jacobian of the synchronization system, that is to be analyzed for stability, is the same as that of the consensus (as the skew symmetric component can be ignored for stability analysis). But the Jacobian, in case of consensus, was guaranteed to be negative-semidefinite as long as all $\nabla^i \mathfrak{grad}^i V_{ij} =\mathfrak{hess}(V)_{g_j^{-1}*g_i}$, were  guaranteed to be positive definite. But since we have assumed that $g_j^{-1}*g_i\in U$ (synchronous solutions in $U$) and hence   $\mathfrak{hess}(V)_{g_j^{-1}*g_i}$ is positive definite for all $i,j$ by Theorem \ref{previs}.  Hence, the Jacobian is negative semi-definite, which by similar techniques, guarantees local exponential stability to the manifold of velocity synchronous solutions, if they exist.

\section{Conclusion and Future Work}
Thus, we have built a unified and a geometrically elegant theory for first and second-order consensus and synchronization, for Lie groups admitting a bi-invariant metric by making use of a special class of functions called $G-$polar Morse functions. We have seen that such functions indeed exist in the commonly encountered groups like $SO(n)$ and $U(n)$. The next interesting question to ask would be, given a compact Lie group, can one always guarantee the existence of such $G-$polar Morse functions? Is a constructive proof possible? This is an interesting mathematical problem. Representation theory ensures that every compact Lie group can be embedded in $U(n)$ and we have a $G$-polar Morse function in $U(n)$ as $V(R)=tr(I-Re[R])$. Can we show that this function, when restricted to any Lie subgroup of $U(n)$, is still a $G$-polar Morse function on the subgroup as well? If the answer is positive, this would give a constructive proof on the existence of such $G$-polar Morse functions on any compact Lie group whose faithful matrix representation by unitary matrices can be constructed. 

Since every compact Lie group has a unitary representation (and hence isomorphic to a subgroup of $U(n)$) and since $V(R)=tr(I-Re[R])$ is a $G$-polar Morse function in $U(n)$, a natural question will be, is the restriction of $V$ on any Lie subgroup $G'$ of $U(n)$, render $V$ as a $G$-polar Morse function on $G'$ as well? The authors would like to explore a definitive answer to this conjecture based on applying Morse theoretic and group theoretic techniques in the future.

The issue of characterizing graphs that have almost globally stable consensus solutions is again a very important open problem. In this work, it was shown to include trees. For oscillators, it is also shown to include complete graphs and sufficiently dense graphs. Extending them to arbitrary compact Lie groups would be a significant contribution.

Other extensions would be to include time-varying adjacency matrices and to extend Theorems \ref{Dorf1},\ref{Dorf} in full generality to the present setting in Lie groups and give the explicit rates for the  exponentials of stability and bounds on gains. The authors also intend to extend these techniques to formation control and tracking on Lie groups with a bi-invariant metric.

\bibliographystyle{plain}      
\bibliography{mainsync}

\begin{thebibliography}{10}

\bibitem{1582620}
V.D. Blondel, J.M. Hendrickx, A.~Olshevsky, and J.N. Tsitsiklis.
\newblock Convergence in multiagent coordination, consensus, and flocking.
\newblock In {\em Proceedings of the 44th IEEE Conference on Decision and
  Control}, pages 2996--3000, 2005.

\bibitem{BulloNS}
F.~Bullo.
\newblock {\em Lectures on Network Systems}.
\newblock Kindle Direct Publishing, {1.6} edition, 2022.

\bibitem{FFBullo}
Francesco Bullo and Andrew Lewis.
\newblock {\em Geometric Control of Mechanical Systems}.
\newblock Springer, New York-Heidelberg-Berlin, 2004.

\bibitem{inbook}
Eduardo Canale and Pablo Monzón.
\newblock Almost {G}lobal {S}ynchronization of {S}ymmetric {K}uramoto coupled
  oscillators.
\newblock 08 2008.

\bibitem{10.2307/2285509}
Morris~H. DeGroot.
\newblock Reaching a consensus.
\newblock {\em Journal of the American Statistical Association},
  69(345):118--121, 1974.

\bibitem{4777127}
Dimos~V. Dimarogonas, Panagiotis Tsiotras, and Kostas~J. Kyriakopoulos.
\newblock Laplacian cooperative attitude control of multiple rigid bodies.
\newblock In {\em 2006 IEEE Conference on Computer Aided Control System Design,
  2006 IEEE International Conference on Control Applications, 2006 IEEE
  International Symposium on Intelligent Control}, pages 3064--3069, 2006.

\bibitem{DORFLER20141539}
Florian Dörfler and Francesco Bullo.
\newblock Synchronization in complex networks of phase oscillators: A survey.
\newblock {\em Automatica}, 50(6):1539--1564, 2014.

\bibitem{HANMANN2006176}
Heinz Hanßmann, Naomi {Ehrich Leonard}, and Troy~R. Smith.
\newblock Symmetry and reduction for coordinated rigid bodies.
\newblock {\em European Journal of Control}, 12(2):176--194, 2006.

\bibitem{kunzz}
Kenneth Hoffman and Ray~A. Kunze.
\newblock {\em {Linear Algebra}}.
\newblock PHI Learning, 2004.

\bibitem{Huy}
C~Huygens.
\newblock Oeuvres complètes de {C}hristiaan {H}uygens.
\newblock 1893.

\bibitem{1205192}
A.~Jadbabaie, Jie Lin, and A.S. Morse.
\newblock Coordination of groups of mobile autonomous agents using nearest
  neighbor rules.
\newblock {\em IEEE Transactions on Automatic Control}, 48(6):988--1001, 2003.

\bibitem{gh}
Philip~Holmes John~Guckenheimer.
\newblock {\em Nonlinear Oscillations, Dynamical Systems, and Bifurcations of
  Vector Fields}.
\newblock Springer New York, NY, {1} edition, 1983.

\bibitem{Milnor}
Milnor J.W.
\newblock {\em Morse Theory}.
\newblock Princeton University Press, 1969.

\bibitem{Khalil}
Hassan~K. Khalil.
\newblock {\em Nonlinear Systems}.
\newblock Prentice Hall, 2002.

\bibitem{kod}
Daniel~E Koditschek and Elon Rimon.
\newblock Robot navigation functions on manifolds with boundary.
\newblock {\em Advances in Applied Mathematics}, 11(4):412--442, 1990.

\bibitem{soa2020first}
Shuyang Ling, Ruitu Xu, and Afonso~S. Bandeira.
\newblock On the landscape of synchronization networks: A perspective from
  nonconvex optimization.
\newblock {\em SIAM Journal on Optimization}, 29(3):1879--1907, 2019.

\bibitem{soa2020}
Jianfeng Lu and Stefan Steinerberger.
\newblock Synchronization of kuramoto oscillators in dense networks.
\newblock {\em Nonlinearity}, 33(11):5905--5918, 2020.

\bibitem{Morse}
Marston Morse.
\newblock The existence of polar non-degenerate functions on differentiable
  manifolds.
\newblock {\em Annals of Mathematics}, 71(2):352--383, 1960.

\bibitem{sarlette2010}
Alain Sarlette, Silvère Bonnabel, and Rodolphe Sepulchre.
\newblock Coordinated motion design on {L}ie groups.
\newblock {\em IEEE Transactions on Automatic Control}, 55(5):1047--1058, 2010.

\bibitem{doi:10.1137/060673400}
Alain Sarlette and Rodolphe Sepulchre.
\newblock Consensus optimization on manifolds.
\newblock {\em SIAM Journal on Control and Optimization}, 48(1):56--76, 2009.

\bibitem{SEPULCHRE201156}
R.~Sepulchre.
\newblock Consensus on nonlinear spaces.
\newblock {\em Annual Reviews in Control}, 35(1):56--64, 2011.

\bibitem{rama-ecc2022}
Rama Seshan, Ravi~N Banavar, and Arun~D Mahindrakar.
\newblock Geometric second-order laplacian flow for consensus on lie groups.
\newblock In {\em 2022 European Control Conference (ECC)}, pages 2191--2195,
  2022.

\bibitem{STROGATZ20001}
Steven~H. Strogatz.
\newblock From {K}uramoto to {C}rawford: {E}xploring the onset of
  synchronization in populations of coupled oscillators.
\newblock {\em Physica D: Nonlinear Phenomena}, 143(1):1--20, 2000.

\bibitem{soa2012}
Richard Taylor.
\newblock There is no non-zero stable fixed point for dense networks in the
  homogeneous {K}uramoto model.
\newblock {\em J. Phys. A: Math. Theor.}, 45(5):055--102, 2012.

\bibitem{rm}
Roberto Tron, Bijan Afsari, and René Vidal.
\newblock Riemannian consensus for manifolds with bounded curvature.
\newblock {\em IEEE Transactions on Automatic Control}, 58(4):921--934, 2013.

\bibitem{ce}
Girvan~M. Wiley~DA, Strogatz~SH.
\newblock The size of the sync basin.
\newblock {\em Chaos: An Interdisciplinary Journal of Nonlinear Science},
  16(1):10.1063--1.2165594, 2005.

\end{thebibliography}

% \subsubsection{Stability  Analysis: first-order Case}

% Evaluating the time derivative of $V_T$ along the trajectories of the first-order system in \eqref{gfolf}, we have 

% \begin{align}
%     \frac{dV_{T}}{dt} = \sum_{k} grad^k V_T (v_k) = \sum_k \mathfrak{grad}^k\hspace{1mm} V_T (v_k)
% \end{align}
%  Substituting for $v_k$ as a negative gradient, we have
%  \begin{align}
%      \frac{dV_{T}}{dt} =  \sum_{k} grad^k V_T (-k_P grad^k V_T) = -k_P ||grad V_T||^2 \leq 0
%  \end{align}
% Substituting  for $V_T$ from \eqref{vt}, we have
% \begin{align}
%     \dot{V}_T =  \frac{1}{2}\sum_k \mathfrak{grad}^k  \sum_{i,j=1}^n V_{ij}   (v_k)
% \end{align}
% Since each $V_{ij}$ is a function of only the positions of agents $i$ and $j$, we have
% \begin{align}
%       \dot{V}_T =  \frac{1}{2} \sum_{i,j=1}^n \mathfrak{grad}^i   V_{ij} (v_i) +  \mathfrak{grad}^j   V_{ij} (v_j) \label{ff}
% \end{align}
% Since $V_{ij}=V_{ji}$, both the terms in the summation at \eqref{ff} above are the same and hence we can write
% \begin{align}
%      \dot{V}_T = \sum_{i,j=1}^n \mathfrak{grad}^i   V_{ij} (v_i)
% \end{align}

% Substituting for $v_i$ from the dynamic \eqref{gfolf}, we have
% \begin{align}
%     \dot{V}_T= \sum_{i,j=1}^n \mathfrak{grad}^i   V_{ij} () 
% \end{align}
%\begin{figure}
%\begin{center}
%\epsfig{file=jcaesar,width=7cm}
%\caption{Gaius Julius Caesar, 100--44 B.C.}
%\label{fig1}
%\end{center}
%\end{figure}

\end{document}